\documentclass[aps,prl,preprint,superscriptaddress]{revtex4-1}

\usepackage{amsmath}
\usepackage[normalem]{ulem}
\usepackage{subfig}
\usepackage{amsfonts}
\usepackage[squaren,mediumqspace,amssymb]{SIunits}
\usepackage{graphicx}

\usepackage{xcolor}

\begin{document}

\title{Use of an electric-dipole forbidden transition to optically probe the Autler Townes effect}


\author{F. Ram\'irez-Mart\'inez}\email[]{ferama@nucleares.unam.mx}\author{F. Ponciano-Ojeda} \author{S. Hern\'andez-G\'omez} 
\affiliation{Instituto de Ciencias Nucleares, UNAM. Circuito Exterior, Ciudad Universitaria, 04510 Ciudad de M\'exico, M\'exico.}
\author{A. Del Angel}\affiliation{Instituto de F\'\i sica, UNAM, Circuito Exterior, Ciudad Universitaria, 04510 Ciudad de M\'exico, M\'exico.}
 \author{C. Mojica-Casique}\author{L.M. Hoyos-Campo}\author{J. Flores-Mijangos}
\affiliation{Instituto de Ciencias Nucleares, UNAM. Circuito Exterior, Ciudad Universitaria, 04510 Ciudad de M\'exico, M\'exico.}
\author{D. Sahag\'un}
\affiliation{Instituto de F\'\i sica, UNAM, Circuito Exterior, Ciudad Universitaria, 04510 Ciudad de M\'exico, M\'exico.}
\author{R. J\'auregui}\email[]{rocio@fisica.unam.mx}\affiliation{Instituto de F\'\i sica, UNAM, Circuito Exterior, Ciudad Universitaria, 04510 Ciudad de M\'exico, M\'exico.}\email[]{rocio@fisica.unam.mx}
\author{J. Jim\'enez-Mier}\email[]{jimenez@nucleares.unam.mx}
\affiliation{Instituto de Ciencias Nucleares, UNAM. Circuito Exterior, Ciudad Universitaria, 04510 Ciudad de M\'exico, M\'exico.}

\date{\today}

\begin{abstract}

We study the Autler Townes (AT) effect derived from a strong electric dipole transition stimulated by a resonant laser beam and probing it by means of a weak electric quadrupole transition with a controlled frequency detuning in a ladder configuration. The experiment was carried out for a $^{87}$Rb atomic gas at room temperature in a velocity-selective scheme. The AT effect was monitored via the splitting of the fluorescence spectra associated with the spontaneous decay to the ground state. The theoretical description incorporates the modification of standard few-level schemes introduced by forbidden electric-dipole transitions selection rules. We develop an analytic ladder three-level scheme to approximate the cyclic $5\mathrm{S}_{1/2} \mathrm{F}=2\rightarrow 5\mathrm{P}_{3/2} F=3 \rightarrow 6\mathrm{P}_{3/2} \mathrm{F}=1,2,3\rightarrow 5 \mathrm{S}_{1/2}\mathrm{F}=2$ path.  Other levels that could have effects on the fluorescence are included via a fourth level with effective parameters. Doppler effects and finite bandwidths of the laser beams are included in the theoretical model to closely reproduce the experimental results.

\end{abstract}


\maketitle


\section{Introduction}
Five decades of mastering the use of laser light have yielded highly sophisticated techniques for preparing quantum states in systems of very different natures. Neutral atoms have been one of the physical scenarios in which a high level of control has been successfully achieved. This is due to the relative simplicity with which it is possible to observe matter-wave coherence at levels ranging from hot atomic samples \cite{Biedermann:2017} to Bose-Einstein condensates (BEC) \cite{Bloch:2008}. A range of sub-Doppler spectroscopy techniques constitute the foundation-stones on which even the most sophisticated quantum manipulations experiments are based \cite{Wieman:1976ge,Preston:1996,Pearman2002,Corwin:1998era}. Desirable advances in these techniques demand sub-megahertz or even higher control resolution  of the atomic states with methods that would preferably minimize collateral modifications to the system. Here we report the realization of two combined atomic-state processes that  deliver a technique fulfilling these conditions for real experimental circumstances.      

On the one hand, effects driven by AC fields coupling atomic states, such as the Hanle effect~\cite{Alnis:2003}, coherent population trapping (CPT)~\cite{Gray:1978jm}, electromagnetically induced transparency (EIT)~\cite{Fleischhauer:2005} or the Autler-Townes splitting (ATS)~\cite{Autler:1955gb,Picque1976,Zhang:2010,Moreno:2019} are potential sources of unprecedented ways to prepare  atomic states because they are consequences of perturbations over quantum systems that may modify their energy levels in rather subtle ways. The ATS, for example, is a manifestation of Stark AC shifts induced by a near resonant optical field that can be neatly controlled in contemporary atomic physics laboratories. The first demonstration of the ATS was reported together with a theoretical description of the phenomenon~\cite{Autler:1955gb}. Further theoretical analysis has enabled its unambiguous distinction from other effects that appear in the same systems such as EIT~\cite{Anisimov:2011fn}, which is originated by Fano interference among different transition pathways. This has motivated a whole series of theoretical and experimental work focused on manufacturing discrimination protocols on specific systems in response to given scientific needs~\cite{AbiSalloum:2010eg,Sun:2014kh, Wang:2015ho}. Alkali atoms have been the preferred physical system for studying these effects because their internal states are easily manipulated and they can even be laser-cooled and trapped.  Researchers have performed experiments for more than two decades motivated by applications ranging from fundamental science to basic technology that could even yield everyday devices in the long term. In experiments with ultracold atoms or BECs, local control of the atomic interactions by using Feshbach resonances has improved its experimental feasibility by tuning them via an AT doublet~\cite{Bauer:2009fqa}. AT doublets have shown to be an excellent ruler to measure the coupling between atoms and the pump beam of Stimulated Raman Adiabatic Passage (STIRAP) processes induced to accurately produce Rydberg states in cold atoms~\cite{Cubel:2005cf}. More recently AT splitting has been demonstrated as a means of light absorption and retrieval on atom-based quantum memories~\cite{Saglamyurek:2018cm}.

On the other hand, atomic state preparation has predominately been performed by the use of transitions which are driven by the first interaction term of their coupling with the electric component of laser field radiation. There is, however, a second term giving rise to less probable transitions that are related to the valence-electron electric quadrupole field. The so-called forbidden transitions have also attracted attention within several research fields. They became interesting to cosmologists for understanding the microwave background due to recombination of hydrogen and helium in early outer space \cite{KAPLAN:1939hw}. Quadrupole ($E2$) transitions have also been useful to shed light over parity violation in fundamental physics due to the asymmetry in absorption and fluorescence spectra that can be explained by exchanges of weak neutral Z0 bosons between the electrons and the nucleus of the atom \cite{Bouchiat:1997kka}. The long-lived states that may be reached via forbidden transitions are attractive for error correction on quantum bits \cite{Langer:2005hn, Preskill:1998gf}. Narrow-dipole lines that can be cleanly excited without strong effects of their neighboring transitions have turned out challenging to spot. A promising alternative is the use of forbidden transitions in lattice-based atomic clocks \cite{Taichenachev:2006is}. This has motivated a number of noteworthy experiments in which $E2$ transitions were observed in laser cooled ions \cite{Rafac:2000gi} and neutral atoms \cite{Bhattacharya:2003io}. 

Nowadays experiments are reaching such subtlety levels that higher degrees of control are demanded in the preparation of atomic states. Further research has been required to complement the first few experimental works done on the absorption spectra of $E2$ transitions of the alkali atoms \cite{Weber:1987ix,Tojo:2004ft,Vadla:2001fr}. Velocity-selective spectroscopy with diode lasers is an ideal framework to do this due to its simplicity.
Our research group has reported a technique capable of resolving hyperfine levels together with their magnetic spin projection in atomic Rubidium at room temperature \cite{PoncianoOjeda:2015cf,PoncianoOjeda:2018fs}. Chan \textit{et al.} developed a similar velocity-selective spectroscopic method to resolve E2 Cesium transitions \cite{Chan:16}. 

\begin{figure}
\begin{tabular}{c c c}
\subfloat[]{\includegraphics[width = 0.3\textwidth]{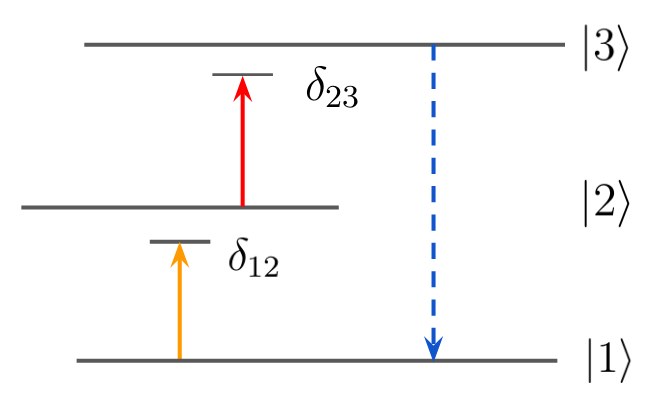}}&
\subfloat[]{\includegraphics[width = 0.36\textwidth]{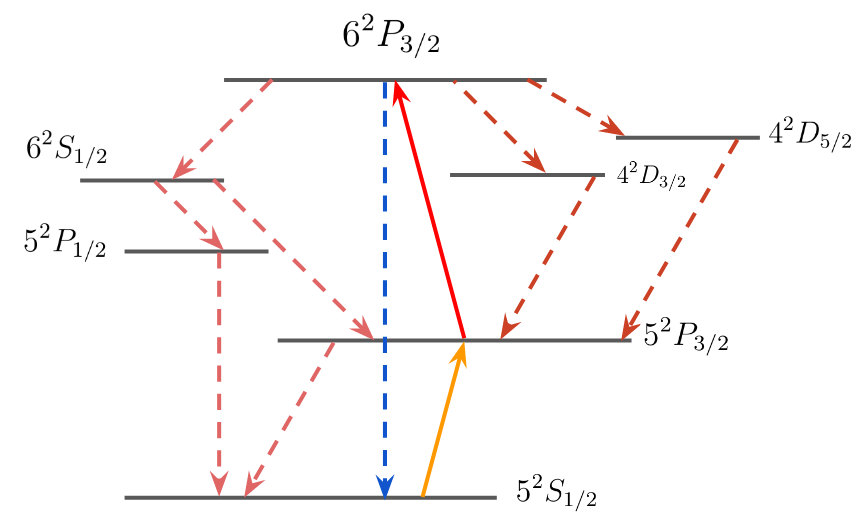}} &
\subfloat[]{\includegraphics[width = 0.3\textwidth]{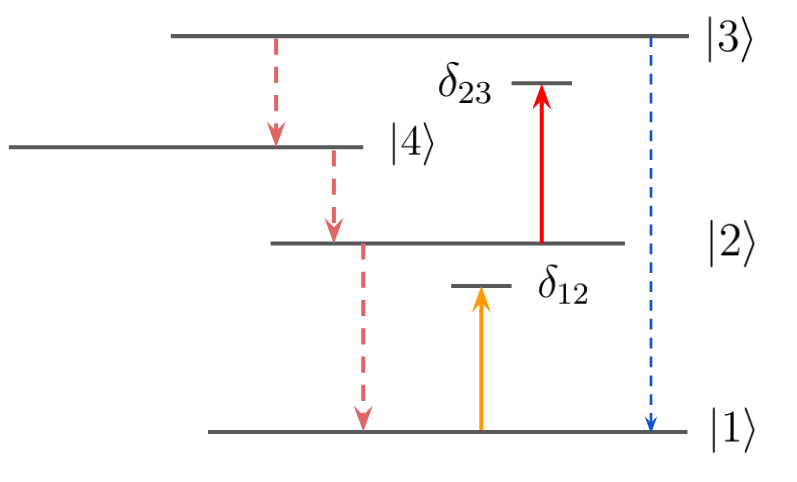}}
\end{tabular}
\caption{(a) Three level diagram showing the stimulated dipole, $\vert 1\rangle\rightarrow \vert 2\rangle$, and quadrupole  $\vert 2\rangle\rightarrow \vert 3\rangle$ transitions as well as the observed spontaneous decay $\vert 3\rangle\rightarrow \vert 1\rangle$. For $^{87}$Rb, $\vert 1\rangle$ is taken as the 5${\mathrm {S}}_{1/2}$, $\vert 2\rangle$ as the $5{\mathrm {P}}_{3/2}$ and $\vert 3\rangle$ as the 6${\mathrm {P}}_{3/2}$ levels; (b) Level diagram showing other relevant transition paths in the actual realization for  $^{87}$Rb  atoms; (c) Simplified four level diagram, in the calculations the fourth level involves the effective decay rates $\Gamma_{34}$ and $\Gamma_{42}$.}\label{fig:Rbenlvl}
\end{figure}

Here we report a detailed experimental and theoretical study of the ATS generated by the strong coupling of an electric dipole transition by probing it with the help of a weak, non-perturbative electric quadrupole transition in a gas of atomic Rubidium at room temperature. To our knowledge there is only one instance in which a dipole forbidden transition is used to probe the ATS; in \citep{Bhattacharya:2003io}, this process is studied in a Sodium magneto-optical trap (MOT) with photoionization detection. In the present work the ATS is non-destructively probed by detecting a fluorescence decay.

In addition, we provide a full theoretical treatment consisting of a master equation model that achieves good agreement with our experiments. A first novelty of the treatment subject to this article is that of a three atomic levels forming a ladder configuration where the coupling field excites state $\vert 1\rangle$ to the $\vert 2\rangle$ state, and the second resonance is excited by a weak probe beam through a forbidden transition [Fig.~\ref{fig:Rbenlvl} (a)]. The atomic structure of Rubidium presents other significant options besides the $\vert 3\rangle \rightarrow \vert 1\rangle$ transition for the state decay path [Fig.~\ref{fig:Rbenlvl} (b)] that exhibit atomic coherence phenomena worthy of focus for further research. However, we found that a model in which a fourth (dump) level is added [Fig.~\ref{fig:Rbenlvl} (c)] fits our experimental results in a rather acceptable way. These three models are described in detail within the theoretical Section of this paper. We include a careful velocity-selective analysis which we found is vital to understand many subtleties observed in the fluorescence spectra. In a similar fashion as in Ref.~\cite{Finkelstein:wc} where a power narrowing mechanism was recently reported in two-color two-photon excitation in ladder systems, our model can resolve subtle experimental parameters like the bandwidth of our excitation lasers which are of a few \mega\hertz. Additionally, we surprisingly found that this ATS may be observed with $90\%$ of the atoms populating the ground state level. Thus we believe that further, improved implementations of this protocol can serve as a sensitive gauge for specific state characteristics for versatile systems such as cold and ultra-cold atomic ensembles. This gauge has minimal perturbing consequences on the atomic system since quadrupole transitions are typically six orders of magnitude weaker than then dipole transitions normally used to induce ATS. 

\section{The physical system}   

\begin{figure}
{\includegraphics[width =0.5\textwidth]{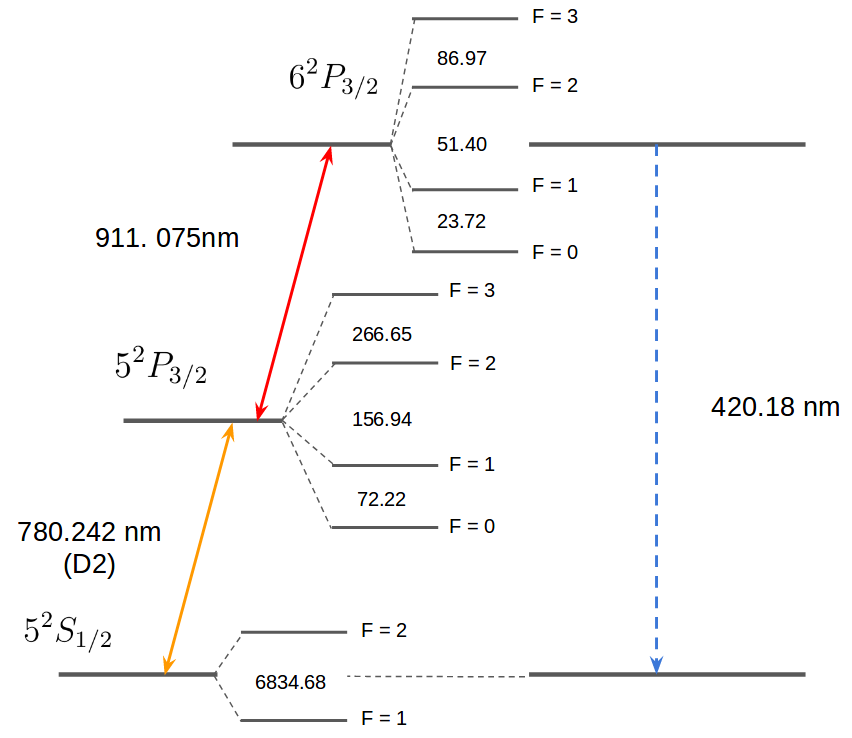}}
      \caption{Hyperfine structure of the $5\mathrm{S}_{1/2}$, $5\mathrm{P}_{3/2}$ and $6\mathrm{P}_{3/2}$ levels of $^{87}$Rb. Numerical values presented in this energy level diagram have been obtained from~\cite{Kurucz:1995}.  \label{fig:AThyper}}
\end{figure}

The measurements presented in the Results and Discussion section of this paper have been performed in a standard room-temperature Rubidium vapor spectroscopy cell. The observations and their theoretical description are reported specifically for $^{87}$Rb, but similar results are obtained for $^{85}$Rb.
Figure \ref{fig:AThyper} shows the set of hyperfine sublevels of the atomic energy structure that is relevant to this work. The numerical values for this energy level diagram, as well as for all the calculations reported throughout the paper, have been either directly obtained or derived from the data reported in by \citet{Kurucz:1995}.
Levels $\left|1\right\rangle $ and $\left|2\right\rangle $ correspond to the ground state $5\mathrm{S}_{1/2}, F=2$ and the first excited state $5\mathrm{P}_{3/2}, F=3$ of the D2 line stretched transition in $^{87}$Rb. These states are resonantly connected with a $\unit{780.242}{\nano\metre}$ laser locked to the transition.
From the $5\mathrm{P}_{3/2}, F=3$ state, level $\left|3\right\rangle $ can only be accessed via the electric quadrupole $E2$ transition. In this case, due to the $E2$ selection rules, either of the $6\mathrm{P}_{3/2}, F=1,2,3$ hyperfine sublevels can be reached with a $\unit{911.075}{\nano\metre}$ laser scanning through this manifold.
The experimental signature of this two-step ladder excitation that is exploited in our measurements is the presence of $\unit{420.18}{\nano\metre}$ fluorescence yield by the $6\mathrm{P}_{3/2}\rightarrow 5\mathrm{S}_{1/2}$ decay.  

In \citet{MojicaCasique:2016iu}, we showed that modeling the evolution of the atomic state populations with Einstein rate equations properly describes the observations. There, by detecting the blue fluorescence, we were able to separately address the $\Delta F=0,\pm 1, \pm 2$ $E2$ selection rules of the $^{87}$Rb $5\mathrm{P}_{3/2} \rightarrow 6\mathrm{P}_{3/2}$ transition by enhancing or suppressing the $\Delta m_F$ with an appropriate choice of the excitation light polarization. Furthermore, in \cite{MojicaCasique:2016iu,PoncianoOjeda:2018fs} it is demonstrated that the choice of the relative  polarization states of the two excitation beams can be utilized to tailor the excitation paths at the level of the magnetic structure of the atoms. Particularly, in the experiments presented on this work a parallel-linear polarization configuration for the preparation and probe beams was employed. This way, $\Delta M_{\parallel}=M_3 -M_2=\pm 1$ transitions are favoured for the ATS probing $E2$ transition. 
This in turn has a direct consequence in the simplification of the atomic structure under study:
the theoretical description of the excitation and decay processes can be independently studied for each of the $\Delta F$ observed for the $E2$ AT probing step.


\section{Theory}
\subsection{Three level atom model. The ideal system.}

The simplest model for describing the general features of the system of interest corresponds to a single three-level atom in the presence of two near-resonant coherent electromagnetic  fields, which is schematically represented in Fig.~\ref{fig:Rbenlvl}(a).  
Note that the configuration under study differs significantly from that described by standard three-level atomic systems. Now, there is a two-photon excitation path that links, via a dipole transition, a ground state level $\vert 1\rangle$ (5S$_{1/2}$, $F=2$ $^{87}$Rb for the experiment reported here) to an excited state $\vert 2\rangle$
(5P$_{3/2}$, $F=3$ for $^{87}$Rb),  followed by a quadrupole
transition to a level $\vert 3\rangle$ (6P$_{3/2}$, $F=1,2,3$ for $^{87}$Rb).  The absorption of two photons in a ladder configuration populates state $\vert 3\rangle$ leading to a spontaneous emission by an electric dipole transition from level $\vert 3\rangle$ to level $\vert 1\rangle$. For $^{87}$Rb, the spontaneous up-conversion process involves two photons of wavelengths $\unit{780}{\nano\metre}$ and $\unit{911}{\nano\metre}$ that yield the emission of a $\unit{420}{\nano\metre}$ photon. In this subsection, we exploit the simplicity of this model to identify some of the general properties of the AT effect under ideal conditions. 
Doppler effects and their partial control through a velocity-selective setup are later included. 

In general, the dynamics of an N-level atom in terms of its density matrix $\rho $ is given by the master equation~\cite{Metcalf} 
\begin{equation}
    \dot\rho_{ll}=\sum_j (\Gamma_{jl}\rho_{jj} - \Gamma_{lj}\rho_{ll}) + i\sum_j (\Omega_{lj}\rho_{jl}
    -\Omega_{jl}\rho_{lj}),\label{eq:mea}
\end{equation}
for the population of level $l$, while for the coherence between levels $n$ and $l$
\begin{equation}
    \dot\rho_{nl} = (i\delta_{nl} - \gamma_{nl})\rho_{nl} +i\sum_j(\Omega_{nj} \rho_{jl} - \Omega_{jl}\rho_{nj}).\label{eq:meb}
\end{equation}
The excitation laser properties are linked to the atomic characteristics via their detuning and the Rabi frequencies. In Eqs.(\ref{eq:mea}-\ref{eq:meb}), $\delta_{nl}$ represents the laser detuning with respect to the transition $n\rightarrow l$ and $\Omega_{nl}$ the corresponding Rabi frequency.
$\Gamma_{mn}$ is the decay rate from level $m$ to level $n$, $\gamma_{mn} = \frac{1}{2} (\Gamma_m + \Gamma_n )$ is the dephasing rate of the coherence $\rho_{mn}$, and $\Gamma_n = 1/\tau_n$  is the total decay rate of level $n$ with $\tau_n$ the lifetime of that level.

For the dipole transition between levels $\left|1\right\rangle $ and $\left|2\right\rangle $ stimulated by a control laser, the corresponding coupling strength $\Omega_{12}$ depends directly on the electric field $\vec E_{c}$ associated with the laser beam of frequency $\omega_c$~\cite{Cohen2011}, 
\begin{equation}
\Omega_{12} =  \frac{\omega_{21}}{\omega_c}\frac{\vec \mu_{21} \cdot \vec E_{c}}{\hbar}, \quad \quad \vec \mu_{21}  = e \langle 2\vert  \vec r \vert 1\rangle,
\end{equation}
where $\omega_{21} =(E_2 - E_1)/\hbar$ is determined by the energy difference between the two levels, $e$ is the electron charge, and $\vec \mu_{21}$ is the electric dipole moment associated to the transition.
 
Levels $\left|2\right\rangle $ and $\left|3\right\rangle $ are weakly connected by the gradient of the electric probe field $\vec E_p$ associated to the laser of frequency $\omega_p$.
If the light field is modeled by an ideal plane wave with wave vector $\vec k_p$, the coupling strength of the quadrupole transition is given by an effective Rabi frequency $\Omega_{23} $~\cite{Freedhoff:1989},
\begin{equation}
\Omega_{23} = \frac{1}{3} \frac{\omega_{32}}{2\omega_p}\frac{\vec k_p \cdot \bar{\bar Q} \cdot \vec E_p}{\hbar}, \quad \quad
\bar{\bar Q}_{ij}   = e \langle 3\vert ( r^2 \delta_{ij}- 3 r_i r_j)\vert 2\rangle
\end{equation}
that involves the atomic quadrupole moment tensor, $\bar{\bar Q}$, and the electric field $\vec E_p$. Since $\vec k_p\cdot \vec E_p =0$, 
\begin{equation}\vec k_p \cdot \bar{\bar Q} \cdot \vec E_p = -e\langle 3 \vert (\vec r\cdot \vec k_p)(\vec r\cdot \vec E_p)\vert 2\rangle.\label{eq:quad}\end{equation}
We shall assume that $\Omega_{12}$ and $\Omega_{23}$ are real numbers.

The Rabi frequency $\Omega_{23}$ of the quadrupole transition is usually small since the typical mean value of the electron distance to the nucleus is much smaller than the wavelength, $k_p r\ll 1$.
A rough estimation shows that for lasers with the same power and for the wavelengths involved in our experimental setup
\begin{equation}\frac{\Omega_{23}}{\Omega_{12}}\sim 10^{-4}.\end{equation}
Besides, levels $\left|2\right\rangle $ and $\left|3\right\rangle $ decay with rates given by $\Gamma_2 $ and $\Gamma_3 $, respectively, and spontaneous decay from level $\vert 3 \rangle$ to $\vert 2 \rangle$ $\Gamma_{32}$ is highly improbable. Thus, $\Gamma_{21} = \Gamma_2 $ and $\Gamma_{31}\sim \Gamma_3 $. 

The master equations for the three level system in the steady state regime ($\dot{\rho}=0 $) can be solved approximately taking into account that $\Omega_{23} \ll \Omega_{12}$. To the lowest order in $\Omega_{23}$ the solutions are:

\begin{eqnarray}
\rho_{11}^{(1)} &=&\frac{4\Omega_{12}^{2}+\Gamma_{21}^{2}+4\delta_{12}^{2}}{
	8\Omega_{12}^{2}+\Gamma_{21}^{2}+4\delta_{12}^{2}}\nonumber\\
\rho_{22}^{(1)} &=& \frac{4\Omega_{12}^{2}}{
	8\Omega_{12}^{2}+\Gamma_{21}^{2}+4\delta_{12}^{2}}\nonumber \\
\rho_{12}^{(1)} &=& \frac{2\Omega_{12}(2\delta_{12}-i\Gamma_{21})}{8\Omega_{12}^{2}+\Gamma_{21}^{2}+4\delta_{12}^{2}}\\
\rho_{13}^{(1)} &=& \frac{4\Omega_{23}\Omega_{12}[4\Omega_{12}^{2}-(  \Gamma_{12}+2i\delta_{12})(\Gamma_{21}+\Gamma_{31}-2i\delta_{23})]}{
	(8\Omega_{12}^{2}+\Gamma_{21}^{2}+4\delta_{12}^{2}) [ 4\Omega_{12}^{2}+    (\Gamma_{21}+\Gamma_{31}-2i\delta_{23})(\Gamma_{31}-2i(\delta_{12}+\delta_{23}) )   ]     }\nonumber\\
\rho_{23}^{(1)} &=& -\frac{8i\Omega_{12}^{2}\Omega_{23}(\Gamma_{21}+\Gamma_{31}-2i\delta_{23})      }{(8\Omega_{12}^{2}+\Gamma_{21}^{2}+4\delta_{12}^{2})
	[4\Omega_{12}^{2}+  (\Gamma_{21}+\Gamma_{31}-2i\delta_{23})(\Gamma_{31}-2i(\delta_{12}+\delta_{23}) )    ]}\nonumber
\end{eqnarray}

\noindent The last element of the density matrix, $\rho_{33} $, is such that
\begin{equation}
\rho_{33}= \frac{2 \Omega_{23}}{\Gamma_{31}} \text{Im}(\rho_{23}). \label{eq:2333}
\end{equation}
So that its lowest order expression requires a second order calculation in $\Omega_{23} $ resulting in
\begin{eqnarray}
\rho_{33}^{(2)}&=&\frac{16 \Omega_{12}^2 \Omega_{23}^2}{\Gamma_{31}}\frac{h(\delta_{23};\Omega_{12},\Gamma_{31},\Gamma_{21})}
{f(\delta_{12};\Omega_{12},\Gamma_{31},\Gamma_{21})g(\delta_{23},\delta_{12};\Omega_{12},\Gamma_{31},\Gamma_{21})} \label{eq:ATprofile}\\
f(\delta_{12};\Omega_{12},\Gamma_{31},\Gamma_{21})& = &4 \delta_{12}^2+\Gamma_{21}^2+8 \Omega_{12}^2 \nonumber \\
g(\delta_{23},\delta_{12};\Omega_{12},\Gamma_{31},\Gamma_{21})& = &
(4\Omega_{12}^2 + \Gamma_{31}(\Gamma_{31} + \Gamma_{21}) - 4(\delta_{12}(\delta_{12} + \delta_{23}))^2\nonumber \\
&+& 4(\delta_{23}\Gamma_{31} +(\delta_{23} + \delta_{12})(\Gamma_{31} + \Gamma_{21}))^2\nonumber \\
h(\delta_{12};\Omega_{12},\Gamma_{31},\Gamma_{21})&=& (4 \delta_{23}^2 + (\Gamma_{21} + \Gamma_{31})^2)\Gamma_{31} +4\Omega_{12}^2(\Gamma_{21} +\Gamma_{31}).
\end{eqnarray}

Notice that for $\delta_{12} = 0$ and $\Omega_{12}>>\Gamma_{21},$ the stimulated transitions $\vert 1\rangle \leftrightarrow\vert 2\rangle$
saturate, and  $\rho^{(1)}_{11}\sim\rho^{(1)}_{22}\sim 1/2$.

In the typical experimental setup, the AT effect is probed by the absorption of the weak   beam. This absorption profile is proportional to the imaginary part of the $\rho_{23} $ density matrix element.
In our experiment we follow a different approach. 
We probe the population of the upper state $\left|3\right\rangle $ by detecting its fluorescence decay into the ground state $\left|1\right\rangle $ that is proportional to the density matrix element $\rho_{33} $, which is in turn also proportional to Im$(\rho_{23}) $, Eq.~(\ref{eq:2333}). As a consequence, the general characteristics of the observed AT effect can be obtained from $\rho_{33}$. 
This effect manifests as the presence of two maxima of $\rho_{33}^{(2)}$ as a function of the detuning of the probe beam.
To show this, we first evaluate the critical points of $\rho_{33}^{(2)}$.  The condition, $$\frac{\partial\rho_{33}^{(2)}}{\partial \delta_{23}}\Bigr\rvert_{\delta_{23}^c} = 0$$
yields a fifth order polynomial in the critical variable $\delta_{23}^c$ which requires a numerical solution.
In the particular case of a zero detuning of the control beam, $\delta_{12} = 0$, this equation is equivalent to an equation with the structure,
\begin{equation}
\delta_{23}^c(a_4(\delta_{23}^c)^4 +a_2(\delta_{23}^c)^2 + a_0) = 0.    
\end{equation}
As a consequence, it is found that a minimum exists at  $\delta_{23} = 0$; there are also two purely imaginary roots and two real roots, given by
\begin{eqnarray}
\delta_{23}^\pm&=& \pm\Big[\frac{1}{4\Gamma_{31}}\Big(2\Omega_{12}(\Gamma_{21} + 2\Gamma_{31})\sqrt{\eta} - (\Gamma_{21} +\Gamma_{31})\eta     \Big)\Big]^{1/2} \\
 \eta &=&4\Omega_{12}^2 + \Gamma_{21}\Gamma_{31} + \Gamma_{31}^2  \nonumber
\end{eqnarray}
From this expression, the critical value of the Rabi frequency $\Omega_{12}^c$ from which the AT doublet would be formed is found to be,
\begin{equation}
   \Omega_{12}^c =\frac{1}{2}\sqrt{\frac{(\Gamma_{21} + \Gamma_{31})^3}{2\Gamma_{21} + 3 \Gamma_{31}}}\simeq \unit{14.45}{\mega\hertz} \label{eq:Occ}
\end{equation}
The requirement of a minimal value of $\Omega_{12}$ for observing the AT effect, and the fact that $\Omega_{12}^c$
is determined by the decay rates of the involved levels, emphasizes its interpretation as an AC Stark effect.

In the limit $\Omega_{12}\gg \Gamma_{21},\Gamma_{31}$, $\rho_{33}^{(2)}$ achieves a saturation value given by
\begin{equation}
\lim_{\Omega_{12}\to \infty}  \rho_{33}^{(2)} = \frac{2\Omega_{23}^2}{\Gamma_{31}(2\Gamma_{31} + \Gamma_{21})}.\label{eq:linrho33}
\end{equation}

\begin{figure}[ht]
\begin{tabular}{c c c}
\subfloat[]{\includegraphics[width = 0.32\textwidth]{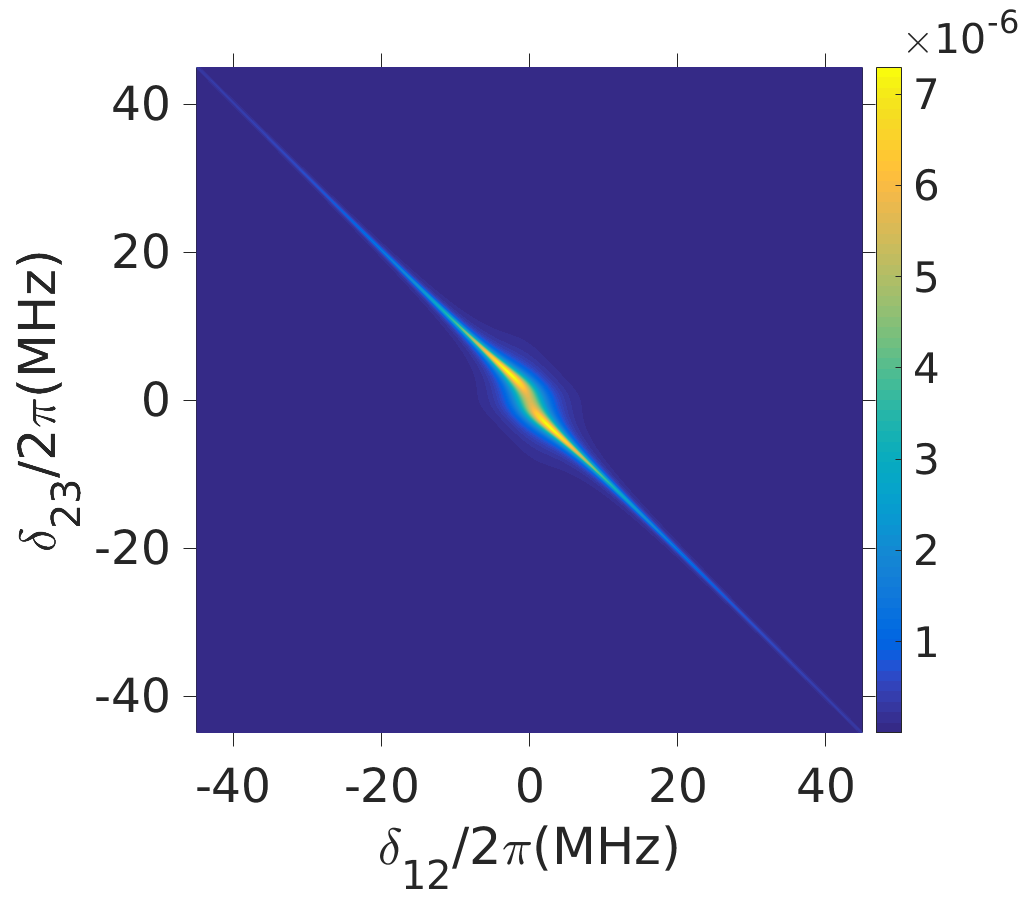}}&
\subfloat[]{\includegraphics[width = 0.32\textwidth]{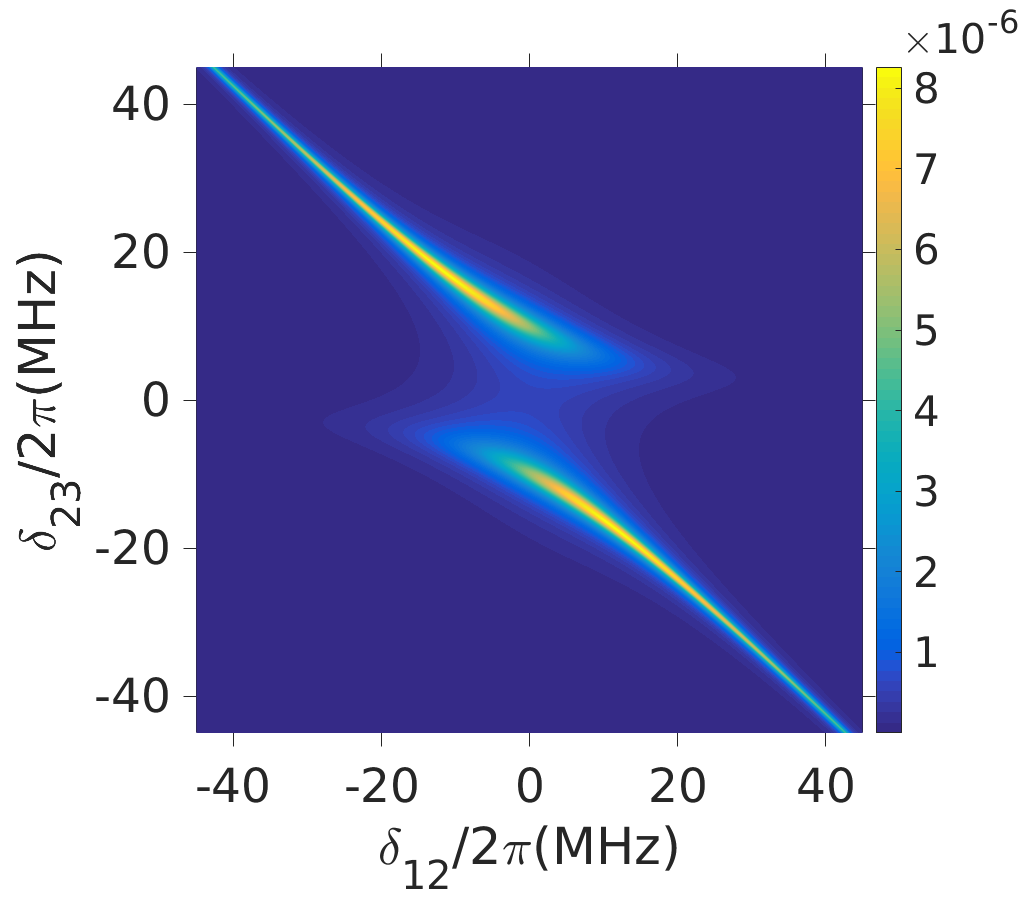}} &
\subfloat[]{\includegraphics[width = 0.32\textwidth]{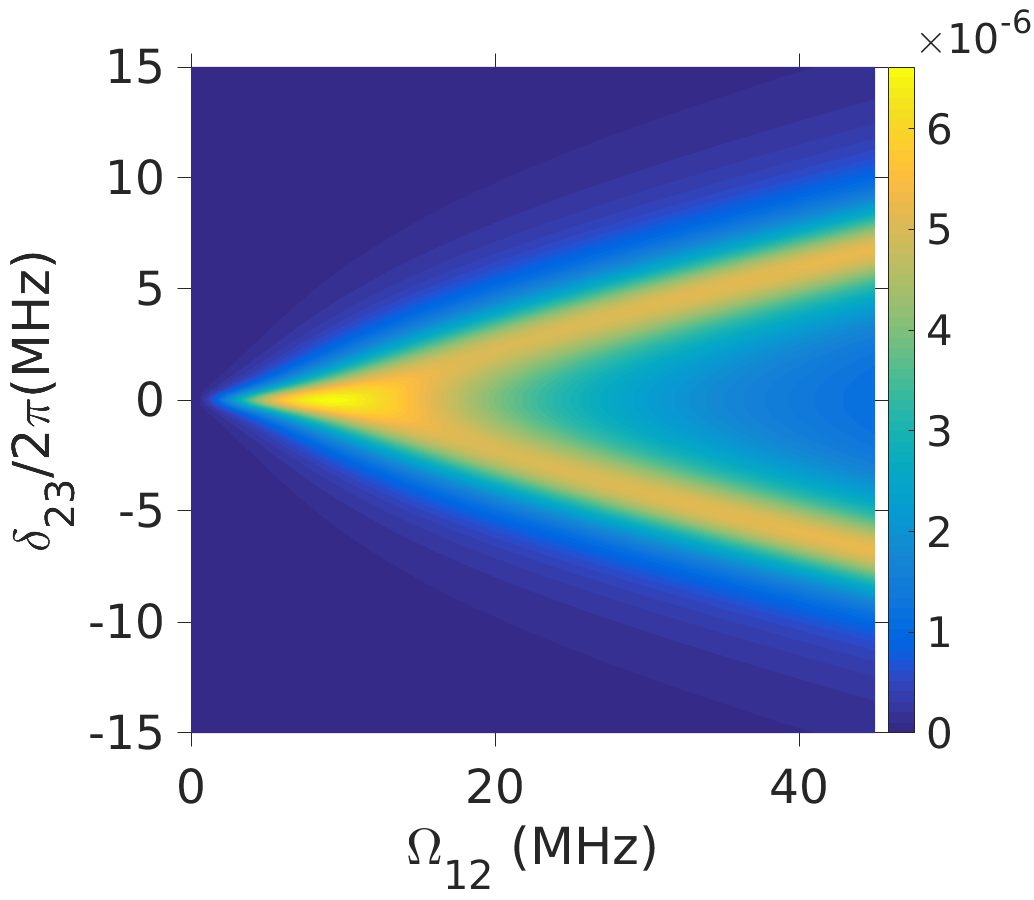}}
\end{tabular}
       \caption{\label{fig:AT:ideal} Ideal steady-state population of the third level as a function of the control laser detuning $\delta_{12}$ and probe laser detuning $\delta_{23}$  (a) at the onset of the AT effect, i.e., $\Omega_{12} =\Omega_{12}^c$, Eq.(\ref{eq:Occ}) and (b) at  $\Omega_{12} = 5\times\Omega_{12}^c$. In (c) the same population is illustrated  as a function of the Rabi frequency $\Omega_{12}$ and the probe laser detuning $\delta_{23}$ taking $\delta_{12}=0$ . The effective Rabi frequency $\Omega^2_{23}$ is taken as 1MHz, which is equivalent to plotting $\rho^{(2)}_{33}/\Omega^2_{23}$.}
\end{figure}

In Fig.~\ref{fig:AT:ideal} the steady-state population of the third level per effective Rabi frequency $\Omega_{23}$ is illustrated using the parameters of the experimental setup. Fig.~\ref{fig:AT:ideal}a shows the dependence with the detunings of both radiation fields with respect to the atomic resonances at the onset of the ATS, $\Omega_{12}^c$.
In Fig.~\ref{fig:AT:ideal}b it can be observed that in the absence of Doppler effects and other dephasing factors, the ATS cannot be obtained by keeping the probe detuning $\delta_{23}$ constant and varying the $\delta_{12}$ control detuning. This happens independently of the power of the control laser. 
Finally, Fig.~\ref{fig:AT:ideal}c demonstrates that the maximum value of $\rho_{33}$ is obtained for  $\Omega_{12}^{max}\sim \unit{9.77}{\mega\hertz}$ which is lower than the critical value $\Omega_{12}^c$ given by eq.~\ref{eq:Occ}. The saturation reached as a function of $\Omega_{12}$ determined by eq.~\ref{eq:linrho33} is also evident in this graph. 

Summarizing, for the ideal model under consideration the general characteristics of the AT effect observed for absorption mediated by dipole electric transitions in a ladder configuration are reproduced for the fluorescence from the top to the bottom level in the unconventional ladder configuration involving a dipole and a dipole forbidden transition. These characteristics include  fluorescence that is symmetric with respect to changes in sign of $\delta_{23}$, a minima at $\delta_{23} = 0$ and just two maxima at $\delta_{23}^\pm$.

\subsection{Doppler effect and velocity selective configuration.}

The model described above does not incorporate several important features that will be present in experiments with room-temperature atomic gases and with realistic laser beams.
Both Doppler effects and finite linewidth of the lasers are two of those properties that could prevent a clear observation of the AT effect. 

It is well known that when a single-frequency laser beam has a frequency $\nu$ which differs slightly from a resonance frequency $\nu_0$ of the atoms in the gas through which it passes, the Doppler shift will make the laser radiation appear exactly on resonance only for those atoms whose component of velocity along the laser beam is $v_z = c(\nu -\nu_0)/\nu_0$. 
Following the same idea, when the control and probe laser beams are used in a counter-propagating configuration, and the control beam frequency is fixed, it is expected that Doppler effects are partially suppressed \cite{Kaminsky1976}. 
The frequency detuning of the laser beams to a transition can be modeled by a Gaussian distribution centered at $\delta_{ij}^{(0)}$ with a spectral width $\sigma_{ij}$,
\begin{equation}
{\mathfrak S}_{ij} (\delta_{ij}) =\frac{1}{\sqrt{2\pi}\sigma_{ij}}e^{-(\delta_{ij} - \delta_{ij}^{(0)})^2/2\sigma^2_{ij}},\quad ij=21,32  .  
\end{equation}
while the Maxwell-Boltzmann distribution describes the probability function of velocities of the atoms at thermal equilibrium at a temperature $T$,
\begin{equation}
{\mathfrak M}(v)=  \Big( \frac{m}{2\pi k_{\mathrm{B}} T}\Big)^{3/2} e^{-m v^2/2k_{\mathrm{B}} T}. 
\end{equation}
An average density matrix that incorporates these detuning effects can be obtained as
\begin{eqnarray}
    \tilde \rho_{ij}^{(\sigma_{lm},T)}(\delta_{21}^{(0)},\delta_{32}^{(0)}) &=&\int\! d ^3 v\int\! d\delta_{21}\int\! d\delta_{32} {\mathfrak S}_{21} (\delta_{21}){\mathfrak S}_{32} (\delta_{32}){\mathfrak M}(\vec v)\rho_{ij}(\delta_{21} + \vec k_{21}\cdot \vec v,\delta_{32} + \vec k_{32}\cdot \vec v)\nonumber\\
    &=& \int\! d ^3 v\int\! d\delta_{21}\int\! d\delta_{32} {\mathfrak S}_{21} (\delta_{21}- \vec k_{21}\cdot \vec v){\mathfrak S}_{32} (\delta_{32}-\vec k_{32}\cdot \vec v){\mathfrak M}(\vec v)\rho_{ij}(\delta_{21} ,\delta_{32})\nonumber \\
\end{eqnarray}
The integration over the velocity can be directly performed. So, for
a counter-propagating lasers configuration, $\hat k_{21} = -\hat k_{32}$,
\begin{equation}
     \tilde \rho_{ij}^{(\sigma_{lm},T)}(\delta_{21}^{(0)},\delta_{32}^{(0)}) =\Big( \frac{m\upsilon_D^2}{k_{\mathrm{B}}T}\Big)^{1/2}
 \int d\delta_{21}\int d\delta_{32} e^{-\kappa(\delta_{21},\delta_{32})}  \rho_{ij}(\delta_{21} ,\delta_{32}) \label{eq:Dwidth}
\end{equation}
where
\begin{eqnarray}
\frac{1}{\upsilon_D^2} &=&\Big(\frac{m}{k_{\mathrm{B}} T} + \frac{\vert k_{21}\vert ^2}{\sigma_{21}^2}  
+ \frac{\vert k_{32}\vert ^2}{\sigma_{32}^2}\Big)\label{eq:vD}\\
\kappa(\delta_{21},\delta_{32})&=& \Big(\frac{\delta_{21}- \delta_{21}^{(0)}}{\sqrt{2}\tilde \sigma_{21}} \Big)^2
 +\Big(\frac{\delta_{32}- \delta_{32}^{(0)}}{\sqrt{2}\tilde \sigma_{32}} \Big)^2\nonumber\\
 &+& \Big(\frac{(\delta_{21}- \delta_{21}^{(0)})}{\sigma_{21}}\frac{\vert k_{21}\vert\upsilon_D}{ \sigma_{21}}\Big)
 \Big(\frac{(\delta_{32}- \delta_{32}^{(0)})}{\sigma_{32}}\frac{\vert k_{32}\vert\upsilon_D}{ \sigma_{32}}\Big)\label{eq:kappa}\\
 \tilde\sigma_{21}^2 &=& \frac{1}{1 - \vert k_{21}\vert^2\upsilon_D^2/\sigma_{21}^2} \sigma_{21}^2\label{eq:sigma21} \\
\tilde\sigma_{32}^2 &=& \frac{1}{1 - \vert k_{32}\vert^2\upsilon_D^2/\sigma_{32}^2}\sigma_{32}^2\label{eq:sigma32} 
\end{eqnarray}
We observe that the Doppler effect, under this ideal counter-propagating configuration, leads both to an
effective increase of the lasers' linewidth by the  factor $\tilde\sigma_{ij}$ and an interference term
that links the two detunings $\delta_{ij}$. A relevant parameter is $\upsilon_D$, Eq.~(\ref{eq:vD}) that has velocity units.

Notice that the more realistic configuration where a spread on the wave vectors $\vec k_{ij}$ is considered can be accomplished by substituting $\delta_{ij}$ by $\delta_{ij} -\vec k_{ij}\cdot \vec v_\bot$, where $\vec v_\bot$ is the component of the velocities of the atoms perpendicular to the main direction of propagation of the laser beams -- the $z$- axis -- followed by the averaging of the resulting expression using the Maxwell-Boltzmann distribution for those components of $\vec v$ and the angular spectra of the laser beams.
This observation is consistent with the expectation that deviations on the realizations of counter-propagating lasers beams alter the AT linewidths \cite{PRA:cp}.
The resulting expression for $\tilde \rho_{ij}^{(\sigma_{lm},T)}(\delta_{21}^{(0)},\delta_{32}^{(0)})$ is given in Appendix A.

\begin{figure}[ht]
\begin{tabular}{c c c}
\subfloat[]{\includegraphics[width = 0.32\textwidth]{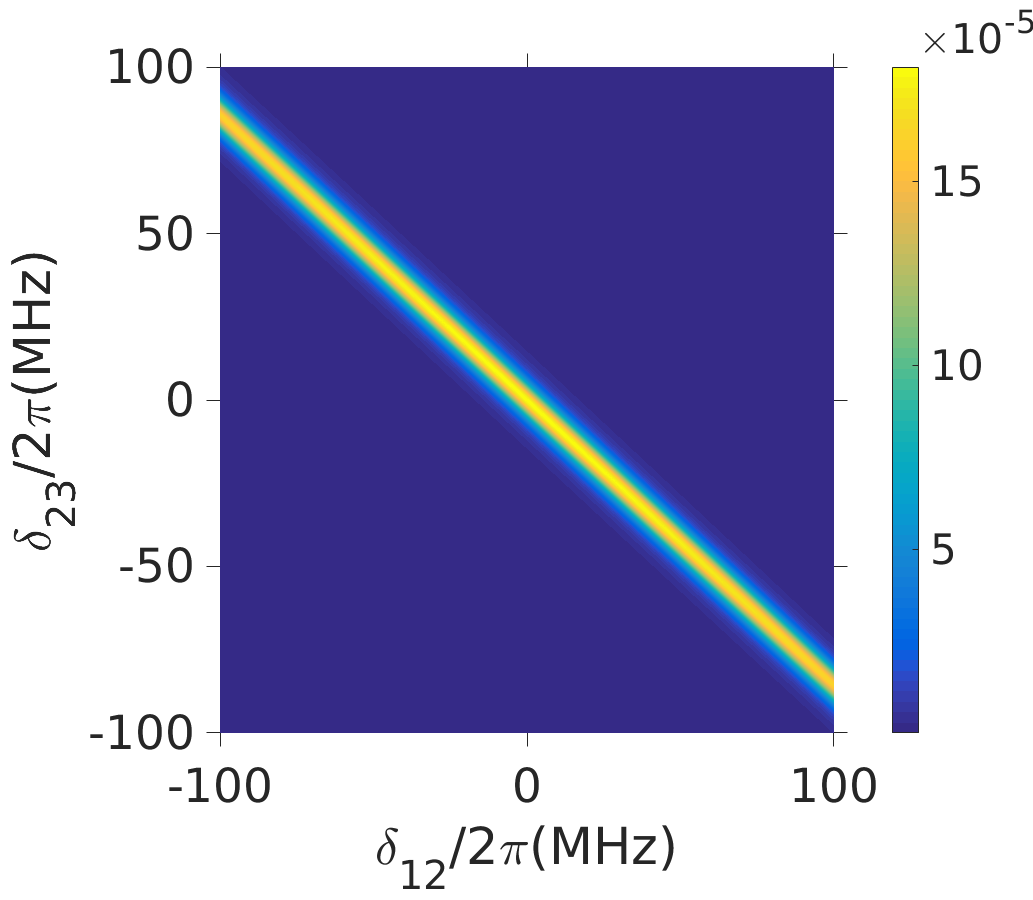}}&
\subfloat[]{\includegraphics[width = 0.32\textwidth]{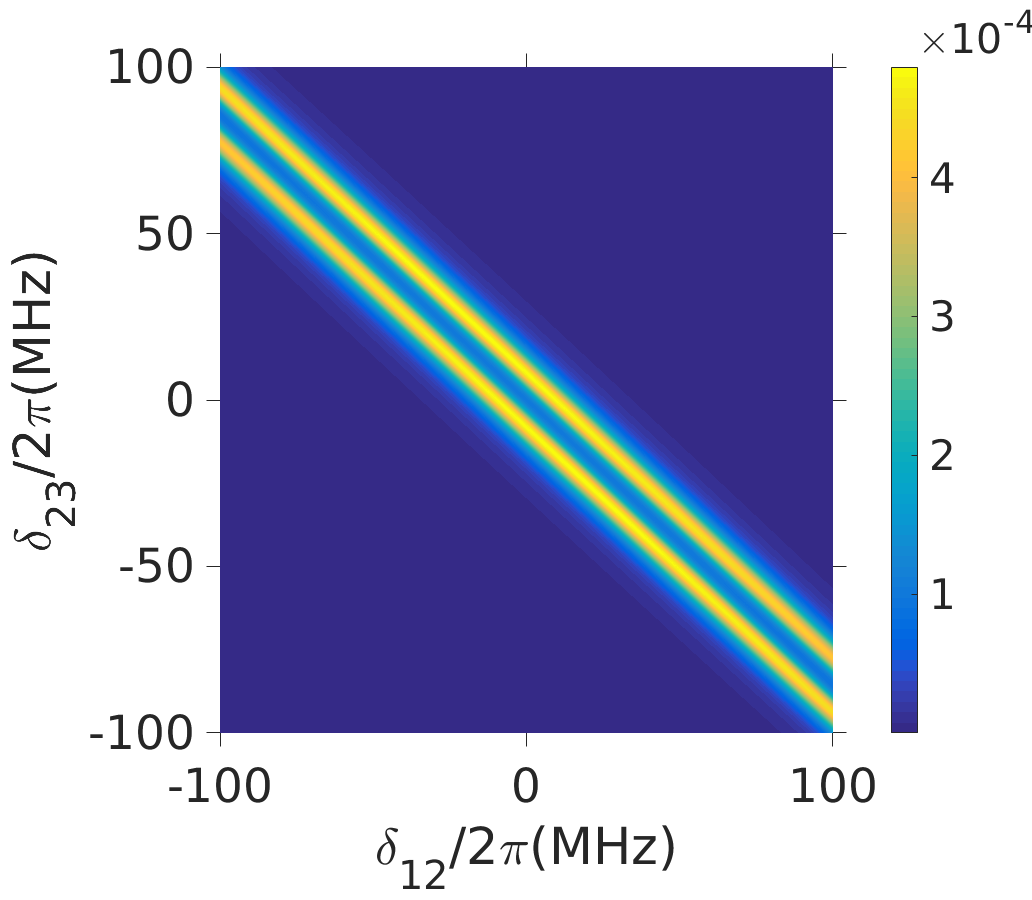}} &
\subfloat[]{\includegraphics[width = 0.32\textwidth]{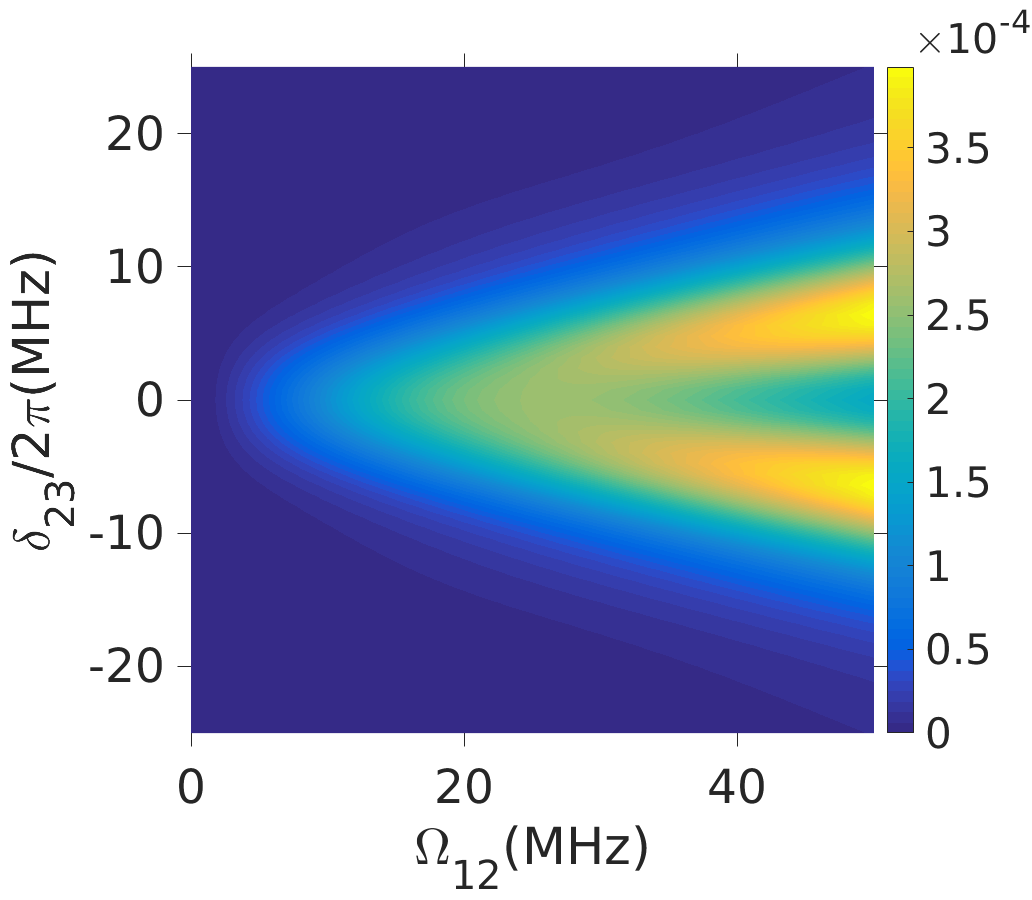}}
\end{tabular}
       \caption{\label{fig:Doppler-3n} Steady-state population of the third level $\tilde \rho_{33}^{(\sigma_{lm},T)}(\delta_{21}^{(0)},\delta_{32}^{(0)})/\Omega_{23}^2$ for an atomic gas at a temperature $T=300^\circ$K in a counter-propagating configuration.
The bandwidth of both control and probe lasers is taken as $\unit{2\pi \times 1.5}{\mega\hertz}$. (a) Displays that variable as a function of the control laser detuning $\delta_{12}$ and probe laser detuning $\delta_{23}$ (a) at the onset of the ATS, i.e., $\Omega_{12} =\Omega_{12}^c$, Eq.(\ref{eq:Occ}) and (b) at  $\Omega_{12} = 5\times\Omega_c^c$. In (c) the same population is illustrated as a function of the Rabi frequency $\Omega_{12}$ and the probe laser detuning $\delta_{23}$ taking $\delta_{12}=0$.}
\end{figure}

The difference in the behavior of the population of the second excited state when including the Doppler contribution can be observed by comparing Figs. \ref{fig:AT:ideal} and \ref{fig:Doppler-3n}.
In Fig.~\ref{fig:Doppler-3n}a and \ref{fig:Doppler-3n}b, the steady-state $\tilde \rho_{33}^{(\sigma_{lm},T)}(\delta_{21}^{(0)},\delta_{32}^{(0)})/\Omega_{23}^2$ is illustrated using the general parameters of the experimental setup and  laser bandwidths $\sigma_{32} = \sigma_{21} =\unit{2\pi \times 1.5}{\mega\hertz}$.
The temperature of the atomic gas is taken as $T=\unit{300}{\kelvin}$, and an ideal counter-propagating lasers configuration is assumed.
Notice that selection of velocities and the finite bandwidth of the lasers allows the observation of the ATS for $\Omega_{12}$ greater than $\Omega_{12}^c$.
As expected, the width of predicted fluorescence profile is larger than in the absence of Doppler and laser finite bandwidth effects.
Besides the interference term in $\kappa(\delta_{21},\delta_{32})$, Eq.~(\ref{eq:kappa}), yields the possibility of observing the AT effect taking the probe (control) detuning as a constant parameter and varying the control (probe) detuning.
In Fig.~\ref{fig:Doppler-3n}c, contrary to its analog in the absence of Doppler and laser bandwidth effects, the maximum of $\tilde \rho_{33}$ is achieved for $\Omega_{12} >\Omega_{12}^c$.
This is a consequence of the selection of velocities that allows that just a small fraction of atoms populate level $\vert 2\rangle$. In fact, the numerical simulations yield $\tilde\rho_{11}^{(\sigma_{lm},T)} >0.9$ for Rabi frequencies $\Omega_{12}<5\Omega_{12}^c$. 
Nevertheless, the order of magnitude of $\rho_{33}$ in the velocity selective scheme is similar or even greater to that expected in the ideal case (no Doppler and zero laser bandwidths).

\subsection{4-level model}
By examining the energy level diagram for $^{87}$Rb, see Fig.~\ref{fig:Rbenlvl}(b), it becomes evident that the decay channels corresponding to transitions 6P${_{3/2}}$ $\rightarrow$ 6S$_{1/2}$, 6P${_{3/2}}$ $\rightarrow$ 4D$_{3/2}$, and 6P${_{3/2}}$ $\rightarrow$ 4D$_{5/2}$ could be significant for the understanding of the characteristics of the $\unit{420}{\nano\metre}$ signal
photons.
Besides this, the 6S and 4D levels also yield a repopulation path to the 5P${_{3/2}}$ level and, in this way, provide a mechanism to modify the occurrence of the quadrupole transition 5P${_{3/2}}$ $\rightarrow$ 6P$_{3/2}$. 

Notice also that in the experimental realization, the laser beams give rise to stimulated transitions only within the region where they have a non-negligible intensity.
For thermal atoms, performing the sequential process of the fast --that is, highly probable-- dipole and slow -- less probable-- quadrupole transition is conditioned by their transit time within the laser beams.
There are cyclic paths that start with a cascade two-photon decay process from the 6P$_{3/2}$ level to either the 6S or any of the 4D states, and then from these states to the 5P$_{3/2}$, followed by an induced quadrupole excitation back to the 6P$_{3/2}$ level.
The achievement of a single cycle is conditioned to atom transit times longer than that required for two slow quadrupole transitions to occur.
Finally, there is also a two photon decay process from the 6S or any of the 4D states to the 5S$_{1/2}$ state.

To model the system, we introduce a fourth level and two effective parameters $\Gamma_{34}$ and $\Gamma_{42}$ that couple it to the other atomic states.
Using the interpretation of the decay rates in terms of a transition probability per unit time, we expect that $\Gamma_{34}$ should be similar to the sum of the decay rates of the 6P${_{3/2}}$ to the 6S$_{1/2}$, 4D$_{3/2}$, and 4D$_{5/2}$ states, that is, if we define
\begin{eqnarray}
    \tilde\Gamma_{34}&=&\Gamma_{6{\mathrm{P}{_{3/2}}}\rightarrow 6\mathrm{S}_{1/2}} + \Gamma_{6{\mathrm{P}{_{3/2}}}\rightarrow 4\mathrm{D}_{3/2}} +\Gamma_{6{\mathrm{P}{_{3/2}}\rightarrow 4\mathrm{D}_{5/2}}}\nonumber\\
    &\sim& (4.506 +0.2346  + 2.11) \mathrm{MHz} = 6.851 \mathrm{MHz},
\end{eqnarray} 
$\Gamma_{34}\sim \tilde \Gamma_{34}$.
Meanwhile $\Gamma_{42}$ is expected to be similar to the weighted average of the decay rates of those states to the 5P$_{3/2}$ levels, that is 
$$\Gamma_{42} \sim $$\begin{eqnarray} 
 (\Gamma_{6{\mathrm{P}{_{3/2}}}\rightarrow 6\mathrm{S}_{1/2}}*\Gamma_{6{\mathrm{S}{_{1/2}}}\rightarrow 5\mathrm{P}_{3/2}} 
 + \Gamma_{6{\mathrm{P}{_{3/2}}}\rightarrow 4\mathrm{D}_{3/2}}&*&\Gamma_{4{\mathrm{D}{_{3/2}}}\rightarrow 5\mathrm{P}_{3/2}} +\Gamma_{6{\mathrm{P}{_{3/2}}\rightarrow 4\mathrm{D}_{5/2}}}*\Gamma_{4{\mathrm{D}{_{5/2}}}\rightarrow 5\mathrm{P}_{3/2}})/\tilde\Gamma_{34}\nonumber\\
&\sim& 12.544  \mathrm{MHz}
\end{eqnarray} 

\begin{figure}[ht]
    \begin{tabular}{c c}
    \subfloat[]{\includegraphics[width = 0.40\textwidth]{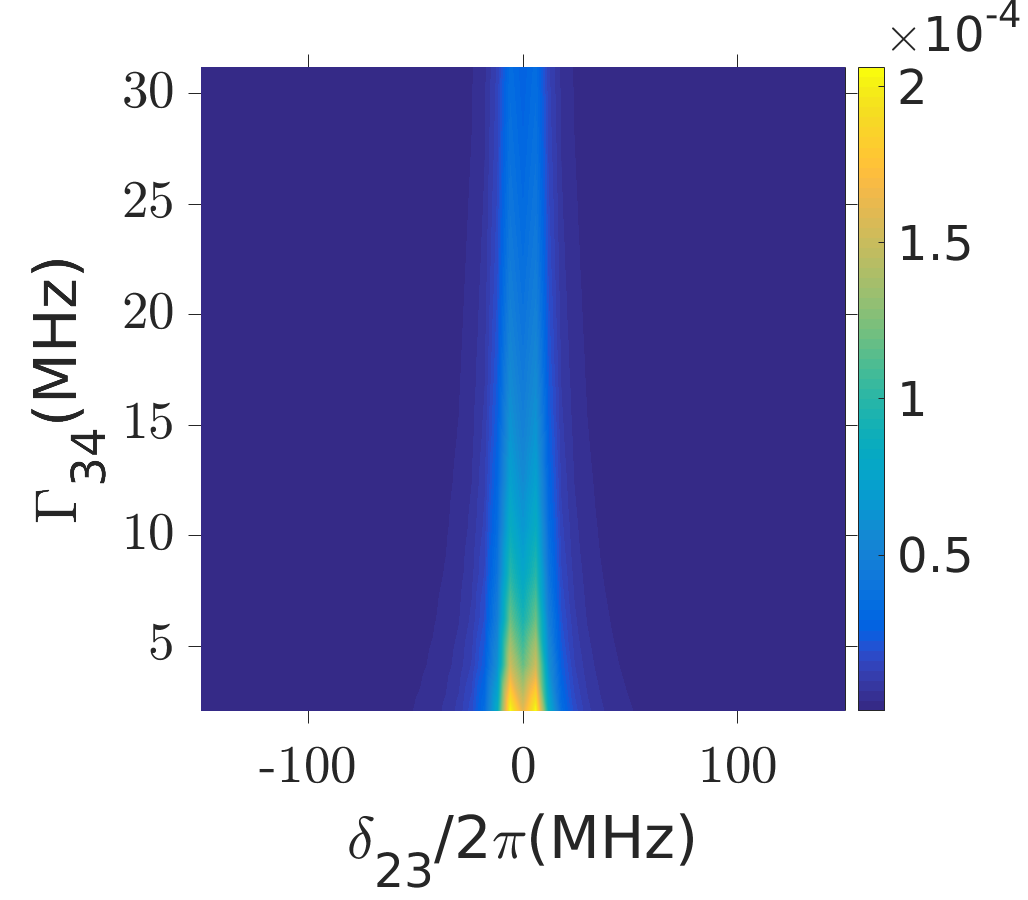}}&
    \subfloat[]{\includegraphics[width = 0.40\textwidth]{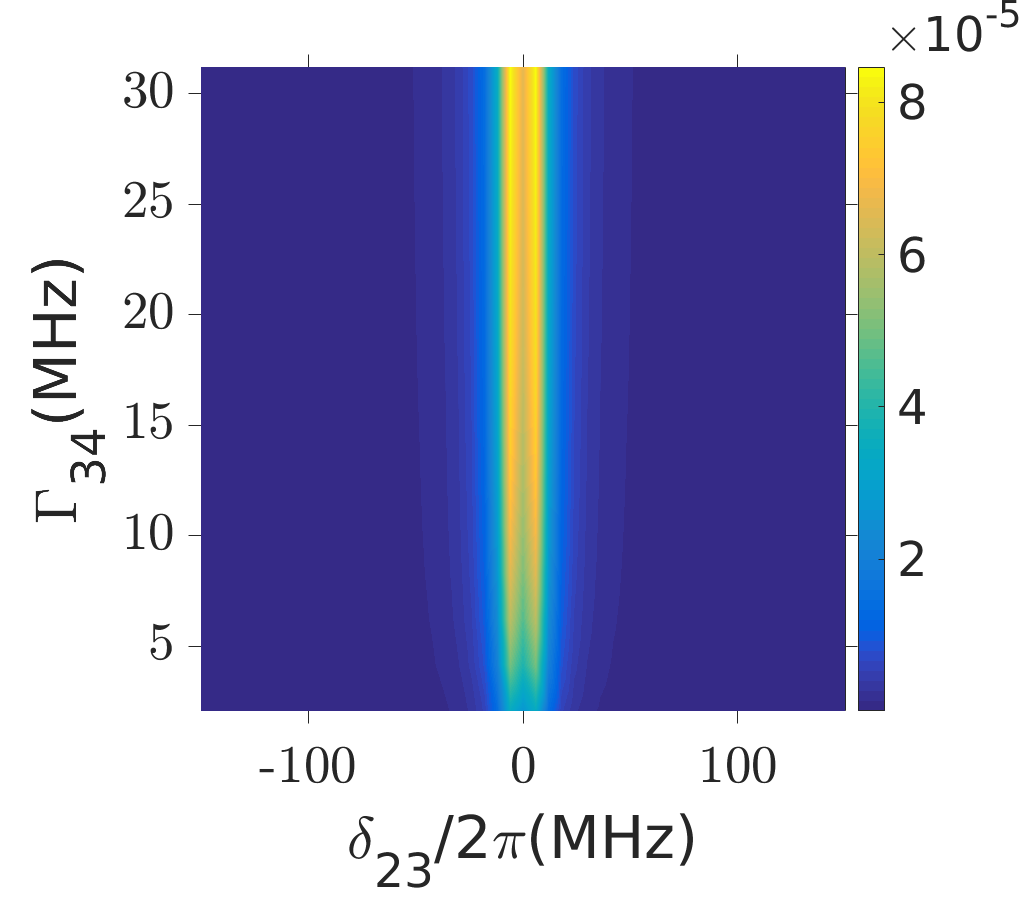}}\\
    \subfloat[]{\includegraphics[width = 0.40\textwidth]{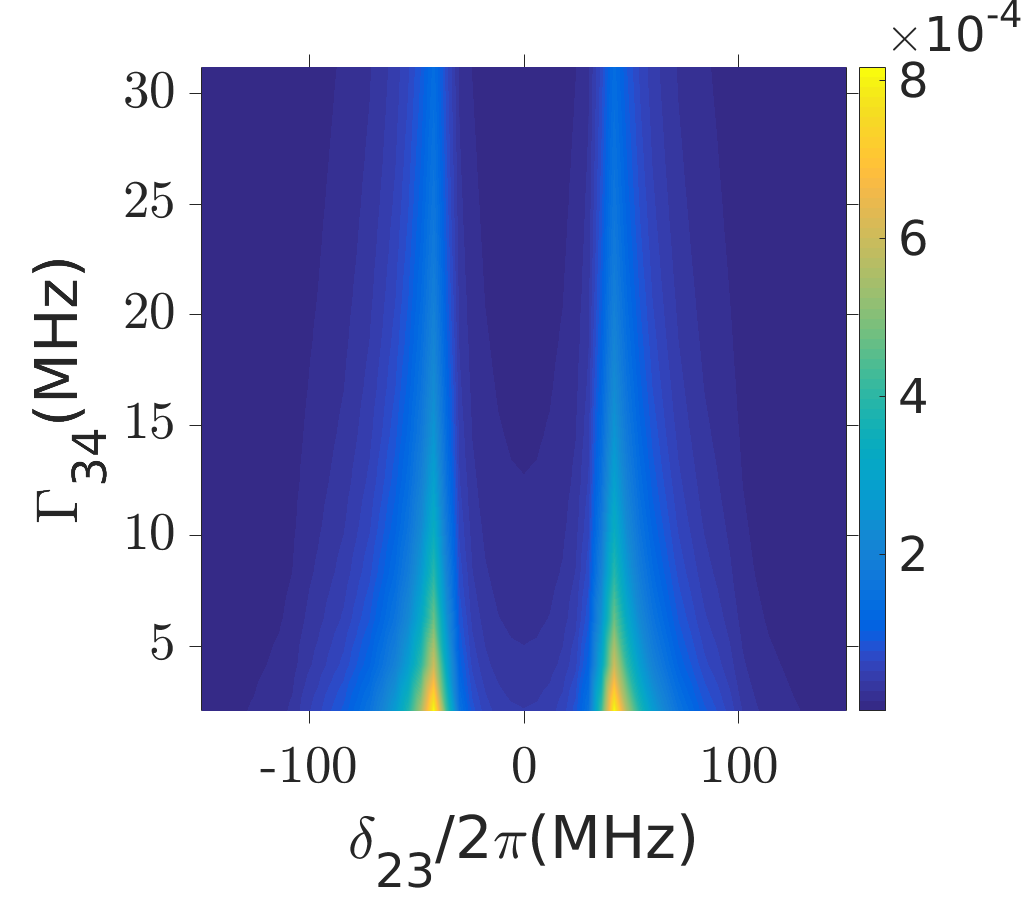}}&
    \subfloat[]{\includegraphics[width = 0.40\textwidth]{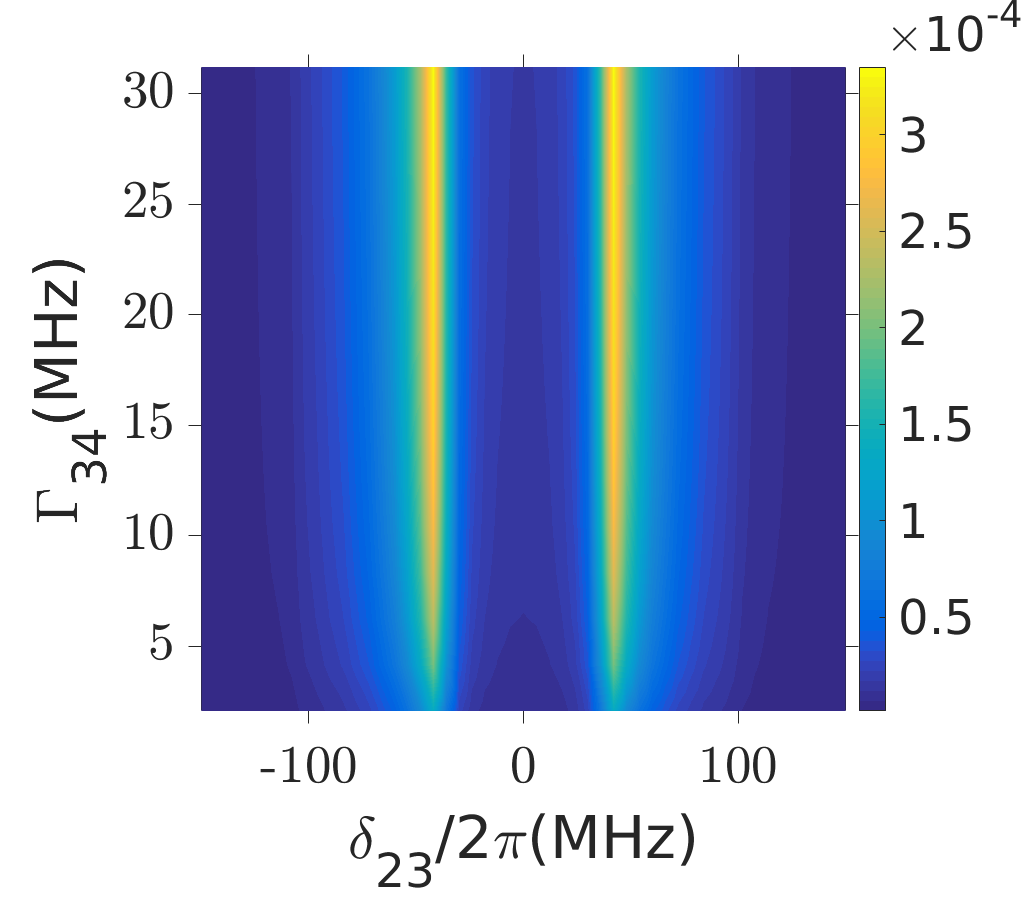}}\\
    \end{tabular}
    \caption{Steady state (a) $\rho_{33}^S=\rho_{33}/\Omega^2_{23}$ and (b) $\rho_{44}^S=\rho_{44}/\Omega^2_{23}$ for  $\Omega_{12} = \unit{15.9}{\mega\hertz}$, and (c) $\rho_{33}^S$ and (d) $\rho_{44}^S$  
    for $\Omega_{12} =\unit{63.07}{\mega\hertz}$. The density matrix elements are illustrated as a function of the detuning $\delta_{23}$ and the decay rate $\Gamma_{34}$, taking $\Gamma_{42} =\unit{12.544}{\mega\hertz}$.
    \label{fig:gammas1}}
\end{figure}

\begin{figure}[ht]
    \begin{tabular}{c c} 
    \subfloat[]{\includegraphics[width = 0.40\textwidth]{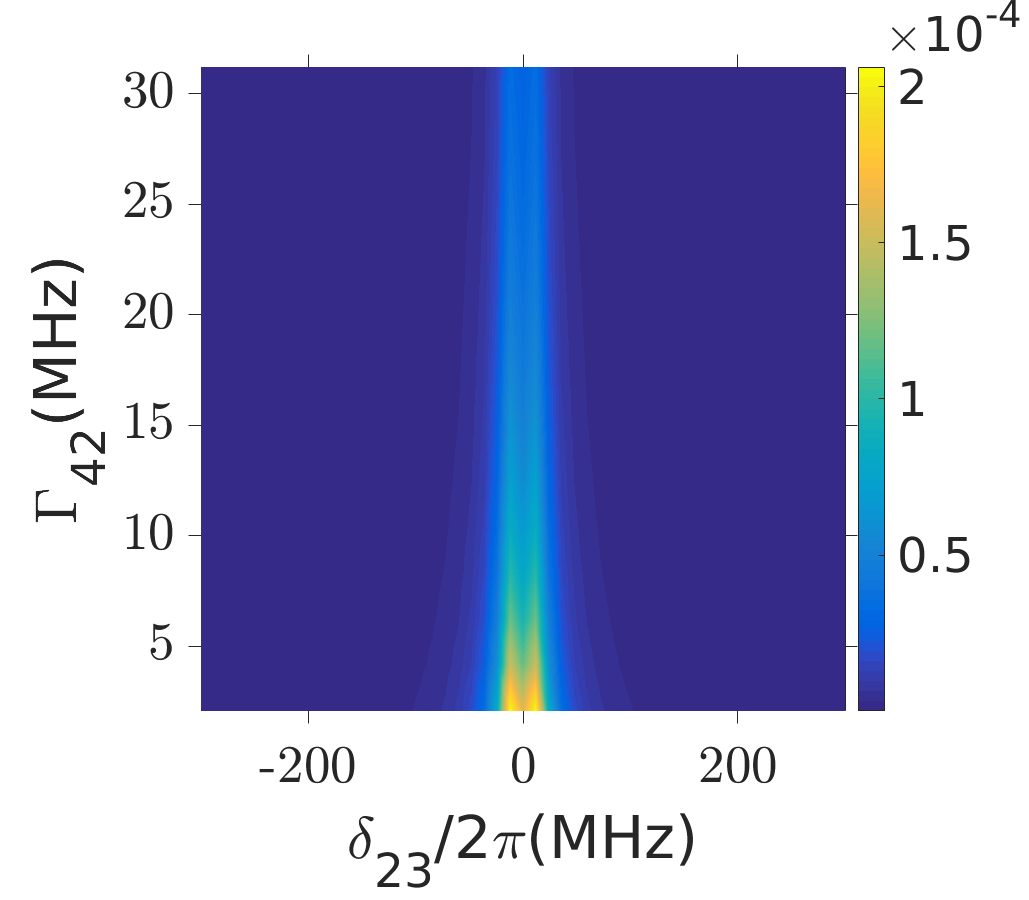}}&
    \subfloat[]{\includegraphics[width = 0.40\textwidth]{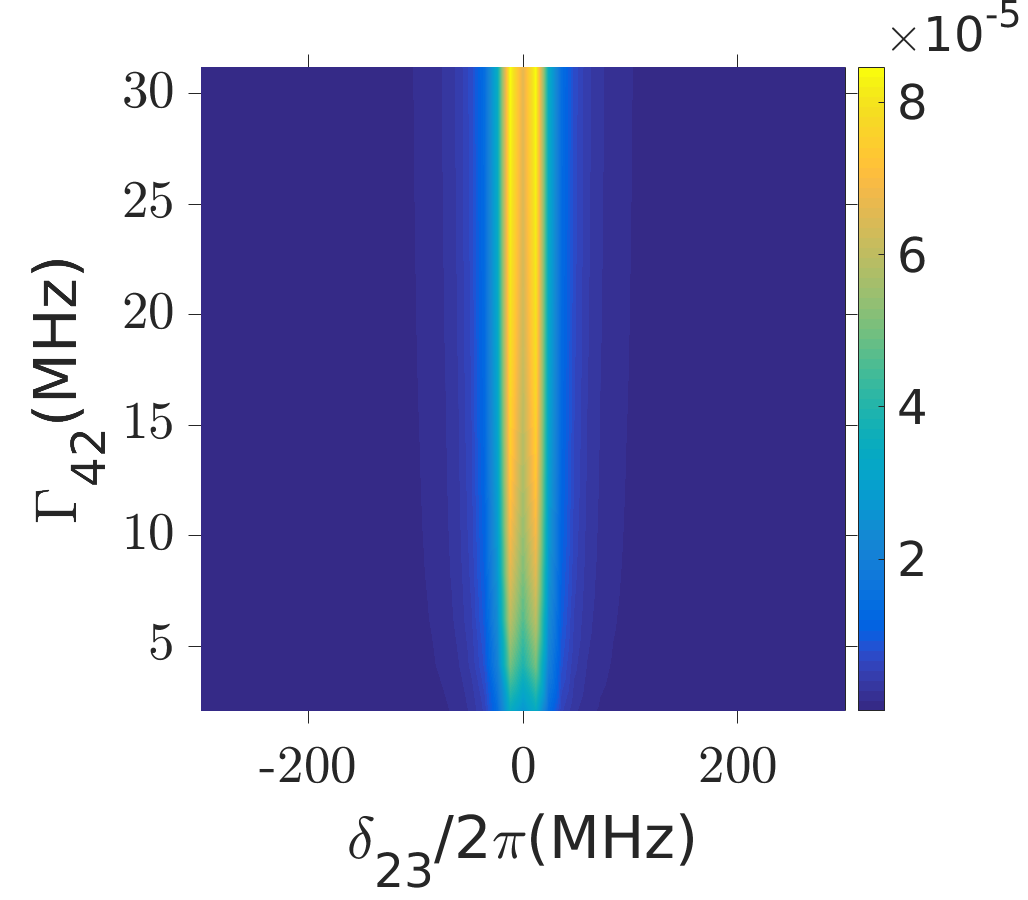}}\\
    \subfloat[]{\includegraphics[width = 0.40\textwidth]{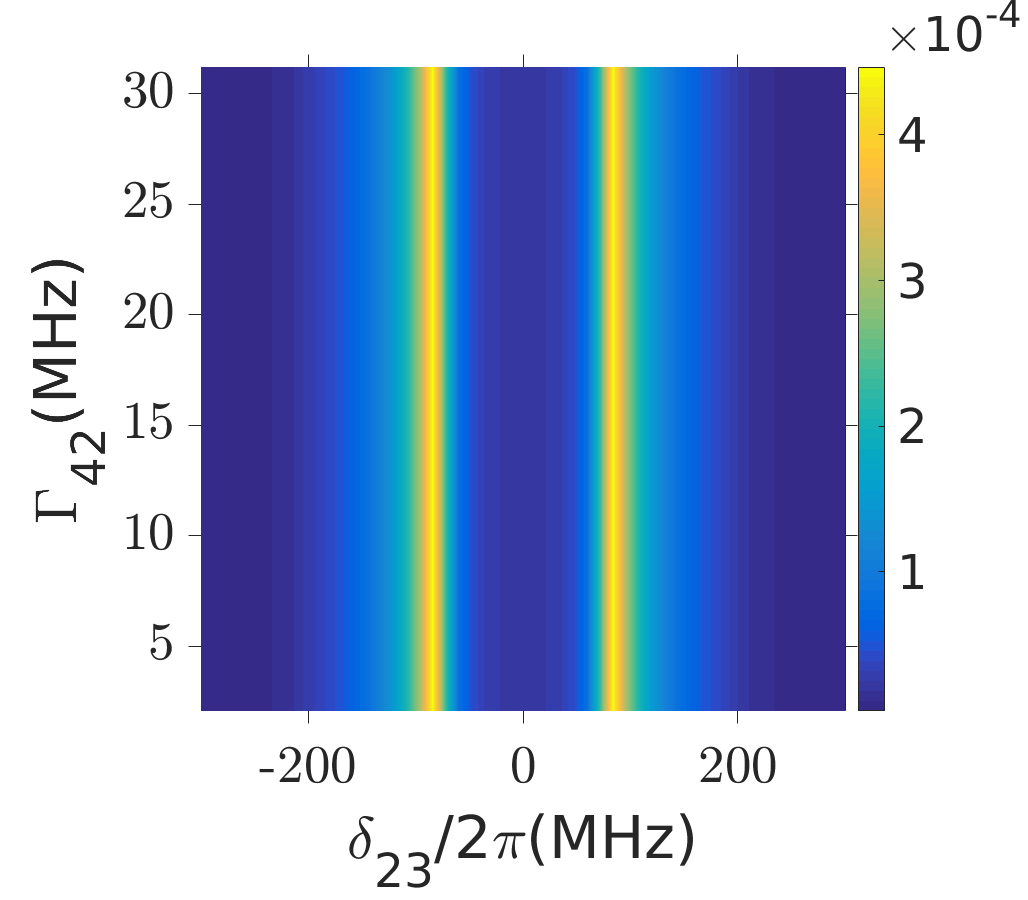}}&
    \subfloat[]{\includegraphics[width = 0.40\textwidth]{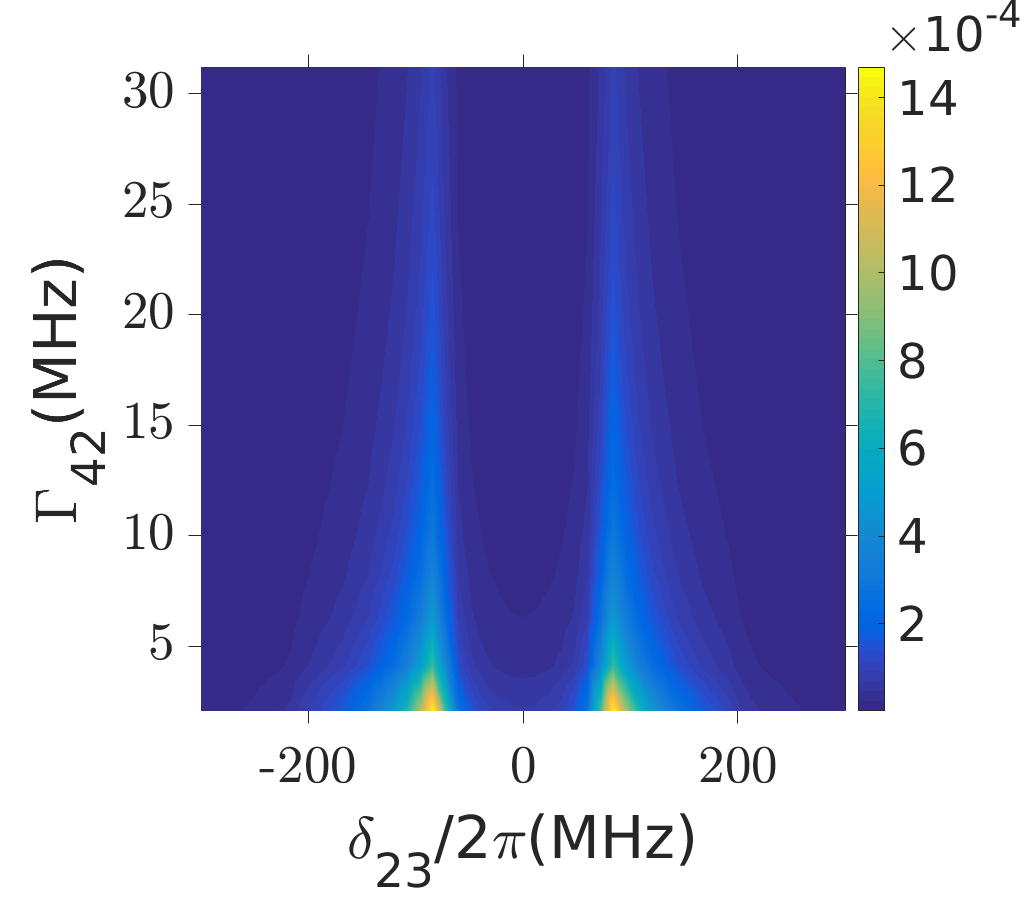}}\\
    \end{tabular}
    \caption{Steady state (a) $\rho_{33}^S=\rho_{33}/\Omega^2_{23}$ and (b) $\rho_{44}^S=\rho_{44}/\Omega^2_{23}$ for  $\Omega_{12} =  \unit{15.9}{\mega\hertz}$, and (c) $\rho_{33}^S$ and (d) $\rho_{44}^S$  
    for $\Omega_{12} = \unit{63.07}{\mega\hertz}$. The density matrix elements are illustrated as a function of the detuning $\delta_{23}$ and the decay rate $\Gamma_{42}$ taking $\Gamma_{34} =\unit{6.851}{\mega\hertz}$.
    \label{fig:gammas2}}
\end{figure}

Numerical calculations were performed to understand the dependence of the velocity-selective scheme on $\Gamma_{34}$ and $\Gamma_{42}$.
This involves solving the time-dependent Bloch equations in the counter-propagating configuration with a velocity dependence on the detunings, and then performing a
velocity average using the Maxwell- Boltzmann distribution.
It was observed that about $\unit{30}{\micro\second}$ are required to achieve a steady solution for the Bloch equations.
This time is slightly higher than that expected for a $^{87}$Rb atom to transit in a transverse path through a Gaussian laser beam with a $\unit{~0.5}{\centi\metre}$ waist at room temperature.
In Figs.~\ref{fig:gammas1}-\ref{fig:gammas2} the results are illustrated for the steady state populations $\rho_{33}$ and $\rho_{44}$ scaled by $\Omega_{23}^2$ to remove the dominant quadratic dependence on that Rabi frequency.
In the numerical simulations we considered $\Omega_{23}=\unit{0.1}{\mega\hertz}$.
Notice that the separation of the AT maxima for a given $\Omega_{12}$ is independent of $\Gamma_{34}$ and $\Gamma_{42}$.
That is not the case for the height and width of the AT peaks which in the case of $\rho_{33}$ are highly dependent on $\Gamma_{34}$, and for $\rho_{44}$ are highly dependent both on $\Gamma_{34}$ and $\Gamma_{42}$.
Notice also that $\rho_{33}$ and $\rho_{44}$ have, in general, the same order of magnitude.

\section{Experimental setup}

\begin{figure}
{\includegraphics[width = 0.75\textwidth]{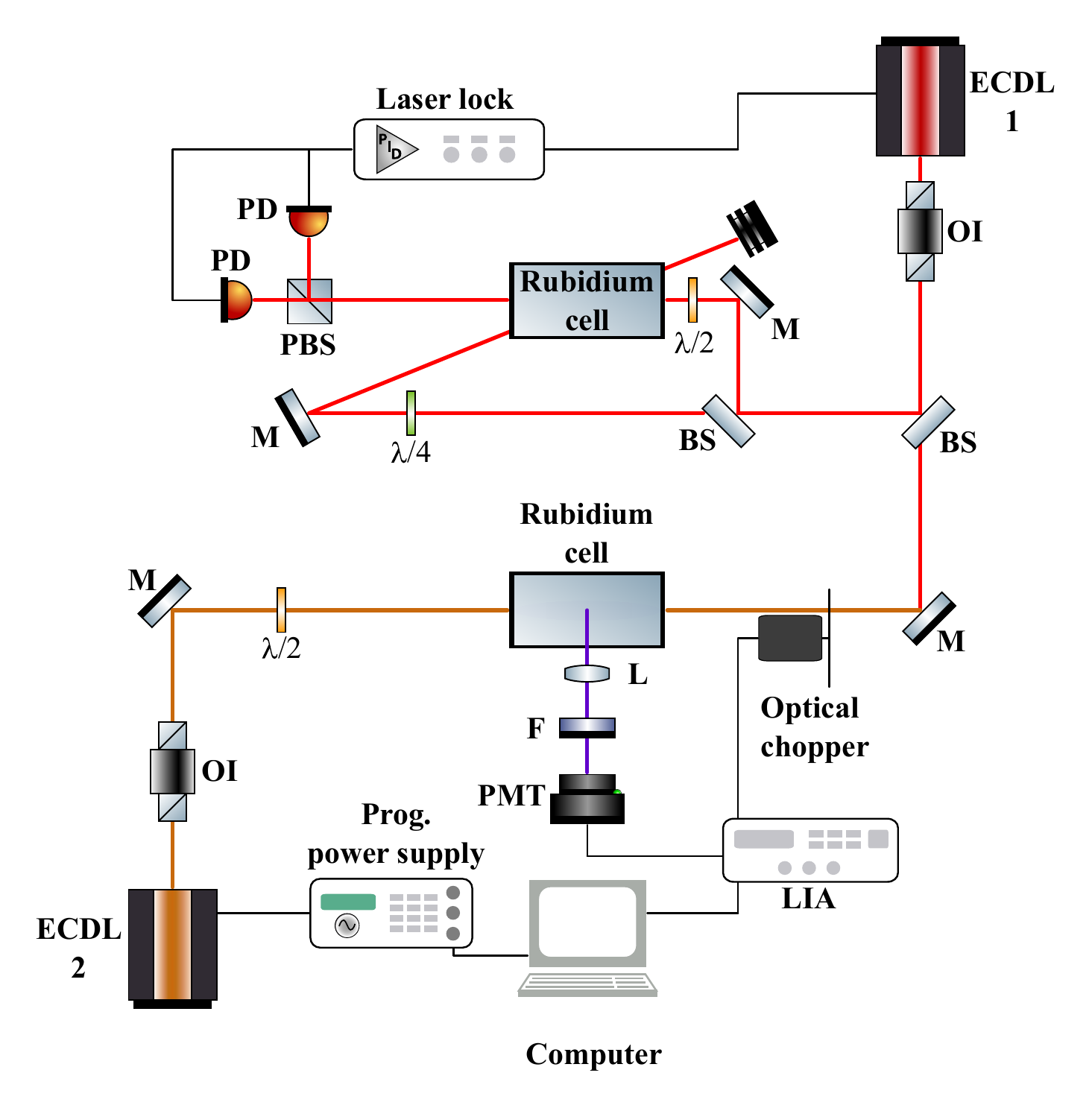}}
       \caption{\label{fig:setup} Experimental setup. ECDL, external cavity diode laser; M, mirror; PMT, photomultiplier tube; BS, beam splitter; PBS, polarizing beam splitter cube; L, lens; F, 420-nm bandpass filter; PD, photodiode; $\lambda/2$, half waveplate; $\lambda/4$, quarter waveplate; OI, optical isolator; LIA, lock-in amplifier.}
\end{figure}

The experimental apparatus has been previously described in detail \citep{PoncianoOjeda:2015cf,PoncianoOjeda:2018fs}.
Figure \ref{fig:setup} is presented here to point out its basic features.
Both the $E1$ and the $E2$ transitions are excited by two home-made extended cavity diode lasers in the Littrow configuration (ECDL1 and ECDL2 respectively).
ECDL1 is stabilized with polarization spectroscopy~\cite{Pearman2002} to the $5\mathrm{S}_{1/2}, F=2 \to 5\mathrm{P}_{3/2}, F=3$, whilst ECDL2 is scanned across the $E2$ manifold to be probed.
Once tuned in frequency, the light from both lasers is polarized and overlapped in a counter-propagating configuration along a room-temperature spectroscopy cell with the natural abundance of the Rubidium isotopes.
The two-step $E1+E2$ excitation is determined by the detection of the $\unit{420.18}{\nano\metre}$ fluorescence that is the result of the $6\mathrm{P}_{3/2}\rightarrow 5\mathrm{S}_{1/2}$ decay.
A fraction of this blue fluorescence light is collected and focused by lens L into a photomultiplier tube (PMT) after passing through a $\unit{420}{\nano\metre}$ interference filter F. This signal is finally amplified by a lock-in system.

The excitation strength of the first step in this two-photon process is given by the Rabi frequency $\Omega_{12}$, which determines the coupling of the atoms with light locked on resonance with the $5\mathrm{S}_{1/2}, F=2 \rightarrow 5\mathrm{P}_{3/2}, F=3$ transition provided by ECDL1. This light pumps atoms from the $\left|1\right\rangle $ state into the 
$\left|2\right\rangle $ state. Then, the state of the atomic system is probed by the much weaker Rabi frequency $\Omega_{23}$ induced by the ECDL2 scanning through the $5\mathrm{P}_{3/2}, F=3 \rightarrow 6\mathrm{P}_{3/2}, F^{\prime}$ manifold. 

To study the forbidden transition leading to the $6\mathrm{P}_{3/2}$ state, it is necessary to have good control of the population of the $5\mathrm{P}_{3/2},F_2M_2$ sub-levels.
Even though it is in fact also a function of the preparation laser intensity, the $M_2$ distribution is mainly determined by its polarization \citep{MojicaCasique:2016iu}.
For reasons that will be explained in the discussion presented in the next section, we chose a parallel-linear polarization configuration for the preparation and probe beams in these experiments.
The preparation laser intensity was always kept below saturation with a power ranging from $\unit{100}{\micro\watt}$ to about $\unit{5}{\milli\watt}$ and an elliptical $\unit{5}{\milli\metre}\times \unit{2.5}{\milli\metre}$ beam profile, while the probe beam power was kept at a power of $\unit{100}{\milli\watt}$ with a $\unit{4.5}{\milli\metre}\times \unit{2.3}{\milli\metre}$ elliptical beam profile.
This corresponds to Rabi frequencies of the control beam $\Omega_{12}$ in the range between $\unit{\sim 20-100}{\mega\hertz}$, calculated as described in the following Section.
The precise evaluation of the probe Rabi frequency $\Omega_{23}$ requires  knowledge of the quadrupole matrix element $\langle 2\vert \bar{\bar Q}\vert 3\rangle$.
A rough estimation, similar to that described in the theory section of this paper and taking into account the probe beam intensity, yields $\Omega_{23}\sim \unit{0.1}{\mega\hertz}$.

\section{Results and discussion.}

\begin{figure}[ht]
\begin{tabular}{c c}
\subfloat[]{\includegraphics[width = 0.48\textwidth]{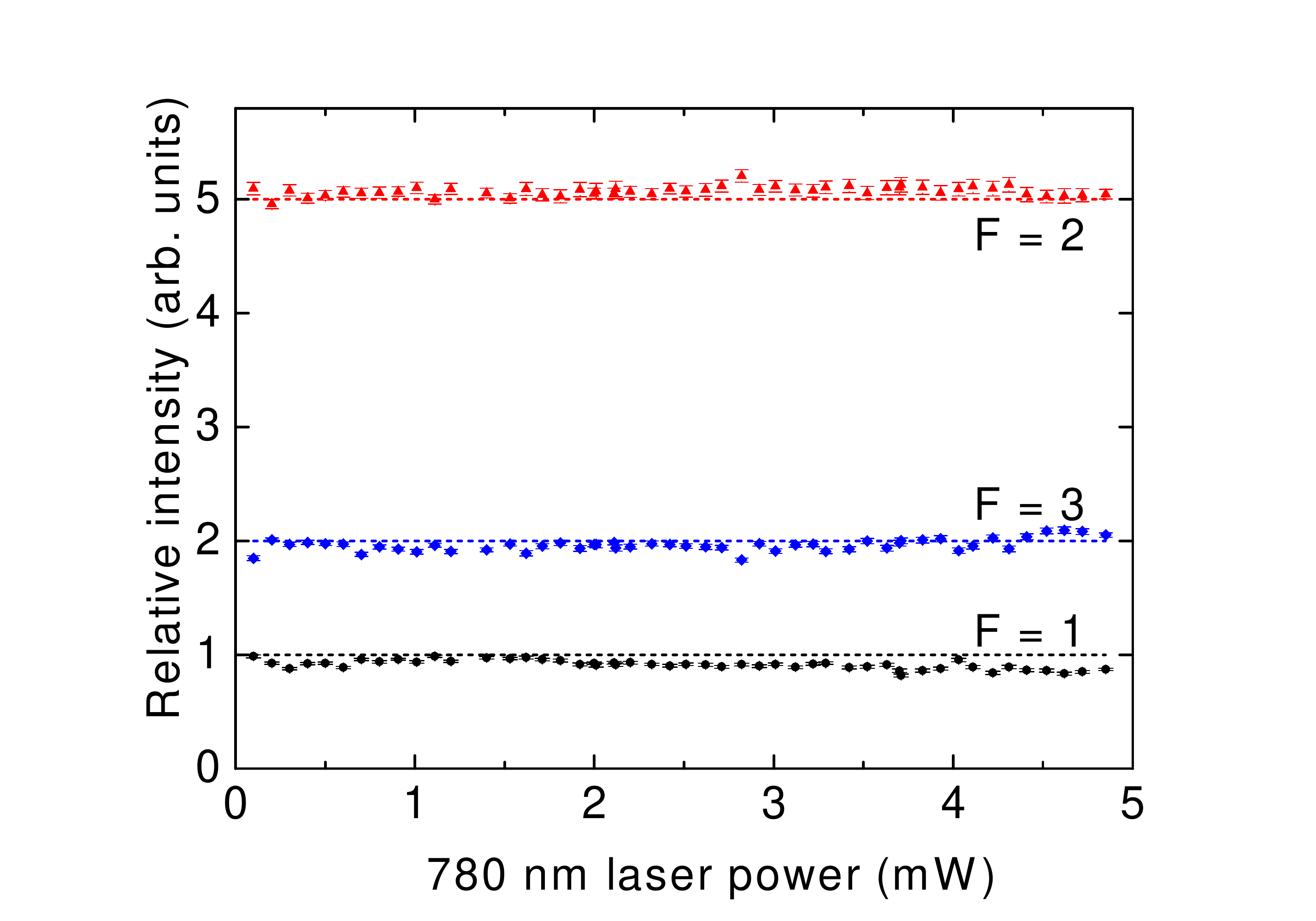}} &
\subfloat[]{\includegraphics[width = 0.48\textwidth]{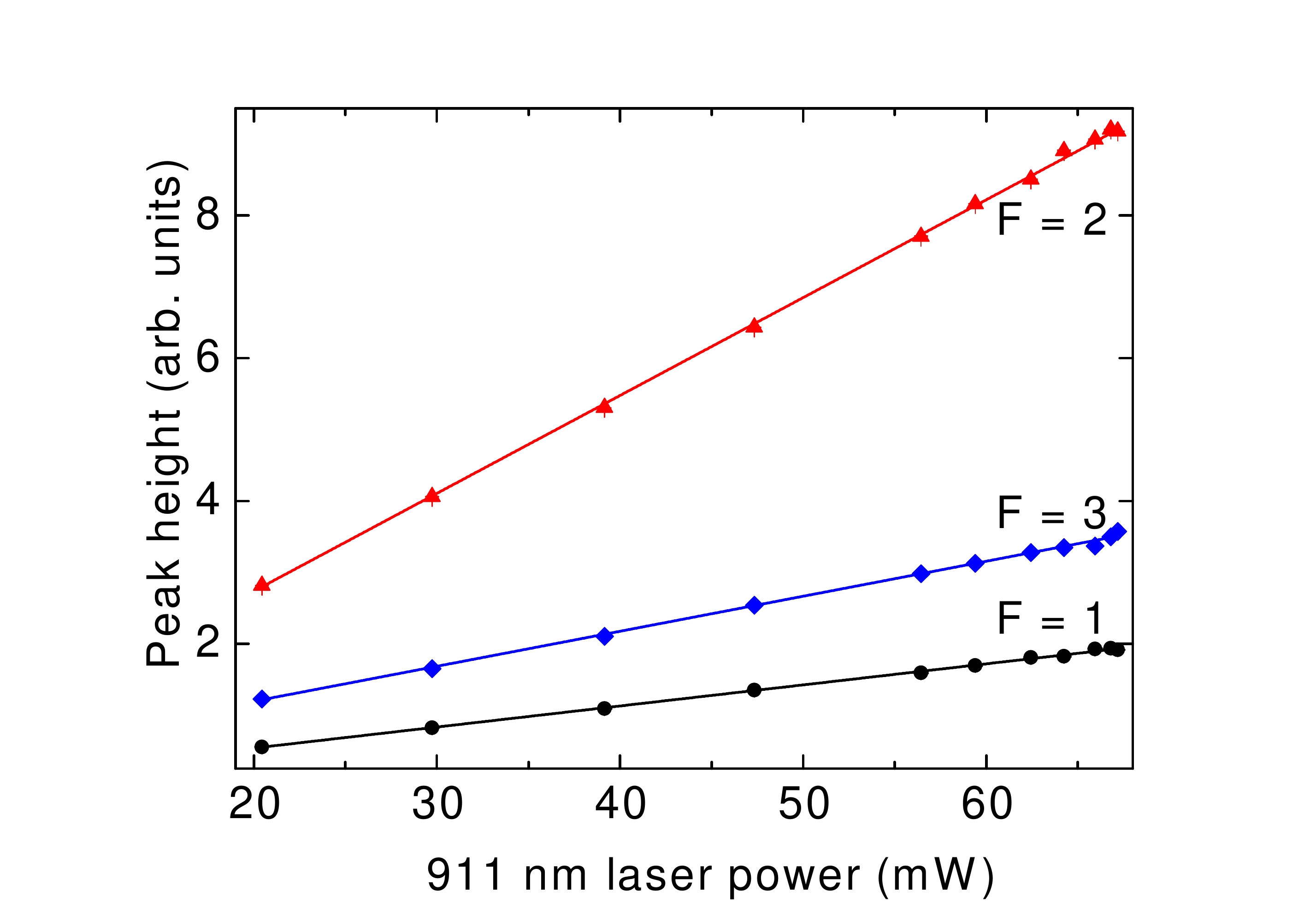}}
\end{tabular}
       \caption{\label{fig:exp} (a) Relative experimental peak height intensities as a function of the power of the $\unit{780}{\nano\metre}$ control beam and (b) the power of the $\unit{911}{\nano\metre}$ probe beam. }
\end{figure}

\begin{figure}[ht]
\includegraphics[width = 1.05\textwidth]{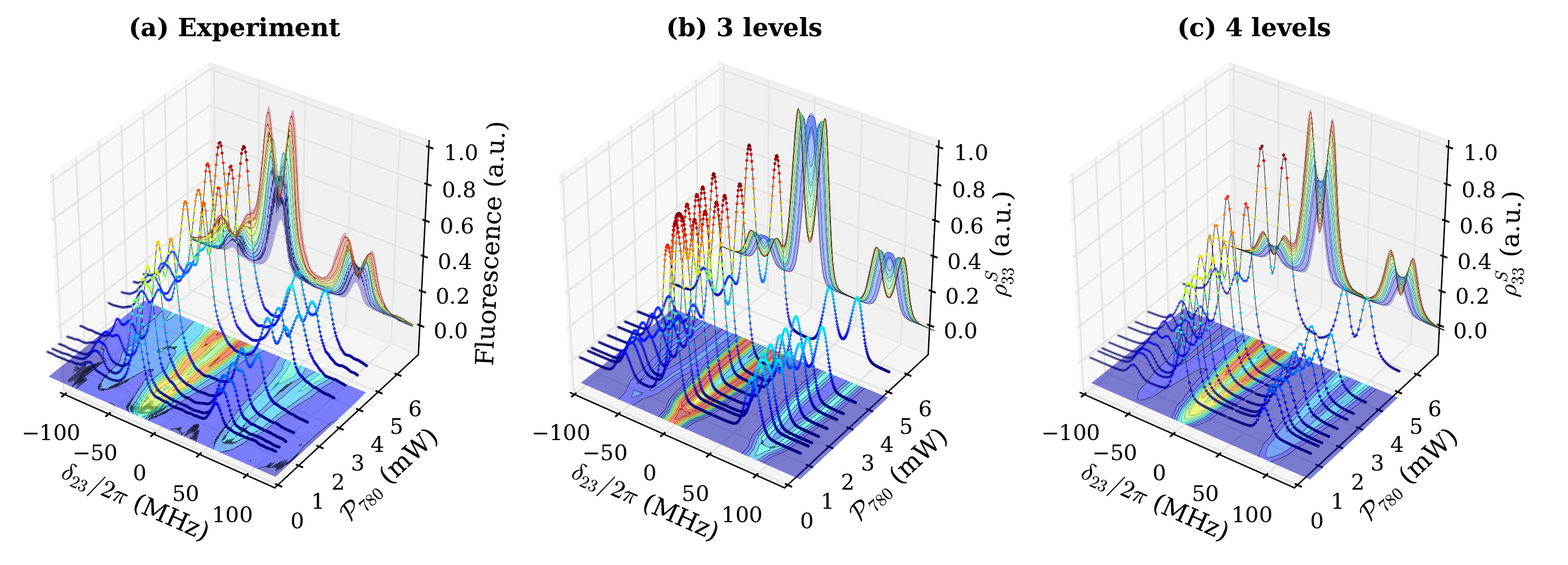}
    \caption{\label{fig:expteo}(a) Illustrative examples of the experimental AT fluorescence of the hyperfine manifold of the 6P$_{3/2}$ state to the 5S$_{1/2}F=2$ level.
    The scale was chosen to facilitate the comparison with (c). Results of the theoretical (b) 3-level and (c) 4-level models for $\rho_{33}^S=\rho_{33}/\Omega_{23}^2$ are shown.
In the theoretical models, an effective bandwidth $\sigma_{21}=\sigma_{32}=\unit{2\pi \times 3.5}{\mega\hertz}$ was used.}
\end{figure}

The choice of the cyclic transition 5S$_{1/2}, F = 2\rightarrow \mathrm{5P}_{3/2}, F  = 3$ stimulated by a linearly polarized laser beam leads to the preparation of a state $\vert 2\rangle$ that can be approximately described by $\mathrm{5P}_{3/2}, F  = 3,\, M_F=0$ with the quantization axis defined by the preparation beam \cite{MojicaCasique:2016iu}. 
As a consequence, the theoretical Rabi frequencies for the $\vert 1\rangle\rightarrow\vert 2 \rangle$ transition can be calculated as \cite{Steck:2008_87}
\begin{equation}
    \Omega_{12}=\frac{eE_c}{\hbar}\frac{1}{\sqrt{5}}\langle 5\mathrm{S}_{1/2}\vert\vert r\vert\vert 5\mathrm{P}_{3/2}\rangle.
\end{equation}
Since the reported experiment involves a quasi counter-propagating linearly polarized $\unit{911}{\nano\metre}$ laser, the relevant quadrupole transition element Eq.~(\ref{eq:quad}) is \cite{MojicaCasique:2016iu}
\begin{equation}
    k_{23} Q_{zx} r_x, 
\end{equation}
taking $\hat e_x$ as the direction of the probe electric field $\vec E_p$.
This leads to the set of independent selection rules for the preparation of the third state: ${\mathrm 6P}_{3/2},F_3=1$, ${\mathrm 6P}_{3/2},F_3=2$, and ${\mathrm 6P}_{3/2},F_3=3$.
One therefore expects to observe three AT profiles centered at the transition frequencies that satisfy the $5\mathrm{P}_{3/2} \rightarrow 6\mathrm{P}_{3/2} $ electric quadrupole selection rules that result in excitation into the $F_3 = 1, 2 $ and $3 $ hyperfine states.
This is illustrated in Fig.~\ref{fig:expteo} where AT profiles are shown for different powers of the control beam.
The relative fluorescence intensities of those profiles are determined by geometric factors involved in the evaluation of the transition matrix elements.
In Refs.~\cite{PoncianoOjeda:2015cf,MojicaCasique:2016iu} it was shown that the probability of observing, due to the two laser excitation, a $\unit{420}{\nano\metre}$ photon resulting from the decay of a given hyperfine state $\vert L_3 J_3 F_3\rangle$ to the state $\vert 5\mathrm \mathrm{S}_{1/2} F_1\rangle$ is given by the expression,
\begin{eqnarray}
    \mathfrak{P}(F_3) &=& \sum_{M_2,M_3,M_{F_1^\prime},\lambda}
    \rho_{22}(F_2,M_2)N\vert\langle 5\mathrm{P}_{3/2}F_2 M_{F_2} \vert Q_{zx}\vert 6\mathrm{P}_{3/2} F_3 M_3\rangle\vert^2 \nonumber \\&\times&
    \vert\langle 6\mathrm{P}_{3/2}F_3 M_{F_3} \vert \mu_{\lambda}\vert 5\mathrm \mathrm{S}_{1/2} F_1 M_{F_1}\rangle\vert^2
\end{eqnarray}
where $\rho_{22}(F_2, M_2)N$ is the population of the $\vert 2\rangle$ state produced by the strong, electric dipole $\vert 1\rangle \rightarrow \vert 2\rangle$
preparation step.
This expression also involves the probabilities of the weak electric quadrupole transition and the electric dipole transition from the 6P$_{3/2}$ hyperfine manifold to the ground state.
A direct calculation yields relative intensities 5:2:1 for $F_3 =2$, $F_3=3$ and $F_3=1$.
This was verified experimentally by comparing the peak height intensities of the measured AT profiles that are shown in Fig.~\ref{fig:exp}a.

The theoretical approach to describe the three AT profiles considered the basic four level model described in Section II for each profile.
We considered the atomic gas at room temperature, assumed that the probe laser is in resonance, and used the effective bandwidths for the two lasers.
The lateral profiles were scaled according to the theoretical proportion 5:2:1 for each $F_3$ state; their central minima corresponded to that given by the hyperfine splitting of $^{87}$Rb \cite{Steck:2008_87}.

The experimental dependence of the peak-height intensities as a functions of the power of the probe beam $\mathcal{P}_{911}$ are shown  Fig.~\ref{fig:exp}(b).
Since $\Omega_{23}^2$ is proportional to $\mathcal{P}_{911}$, the linear dependence on $\mathcal{P}_{911}$ supports the theoretical prediction of the three-level model according to which the population of the $6\mathrm{P}_{3/2}$ level depends quadratically on $\Omega_{23}$.
As a consequence, as mentioned before, $\rho_{33}^S=\rho_{33}/\Omega_{23}^2$ should be quasi-independent of the specific value of  $\Omega_{23}$ used for the evaluation of $\rho_{33}$.
Taking this into account, the theoretical values reported in this Section for $\rho_{33}$ are also scaled by the value of $\Omega_{23}^2$.  

Illustrative examples of the experimental AT fluorescence of the 6P$_{3/2},\, F = 1,2,3$ to the 5S$_{1/2},\, F=2$  manifold are shown in Fig.~\ref{fig:expteo}.
The measured fluorescence as a relative variable has arbitrary units that have been chosen to facilitate its comparison with the theoretical expectations that are given in terms of $\rho_{33}^S$.
In the theoretical results, effective bandwidths of the exciting beams $\sigma_{21}$ and $\sigma_{32}$ were also
optimized to reproduce the experimental data for the central AT profile, and they are the only free parameters.
Their resulting value was greater than the experimental laser bandwidths ($\unit{2\pi \times (1.34 \pm 0.01)}{\mega\hertz}$) reflecting other expected broadening line effects.
Among them are the non-perfect counter-propagating configuration of the control and probe lasers, as well as the distribution of atomic transit times within the laser profiles of the atoms in the Rb gas.
The results of taking $\sigma_{21} = \sigma_{32}= \unit{2\pi \times 3.5}{\mega\hertz}$ are illustrated in Fig.~\ref{fig:expteo}(b) and Fig.~\ref{fig:expteo}(c) for the three- and four-level models respectively.
Notice that the numerical simulations taking into account the velocity selective scheme predict $\rho_{22}(F_2, M_2)< 0.015$ for both models, which is far below saturation.
Nevertheless, the order of magnitude of $\rho_{33}^S$ can be comparable and even greater than that given by the zero temperature three-level model.

\begin{figure}[ht]
\includegraphics[width = 0.6\linewidth]{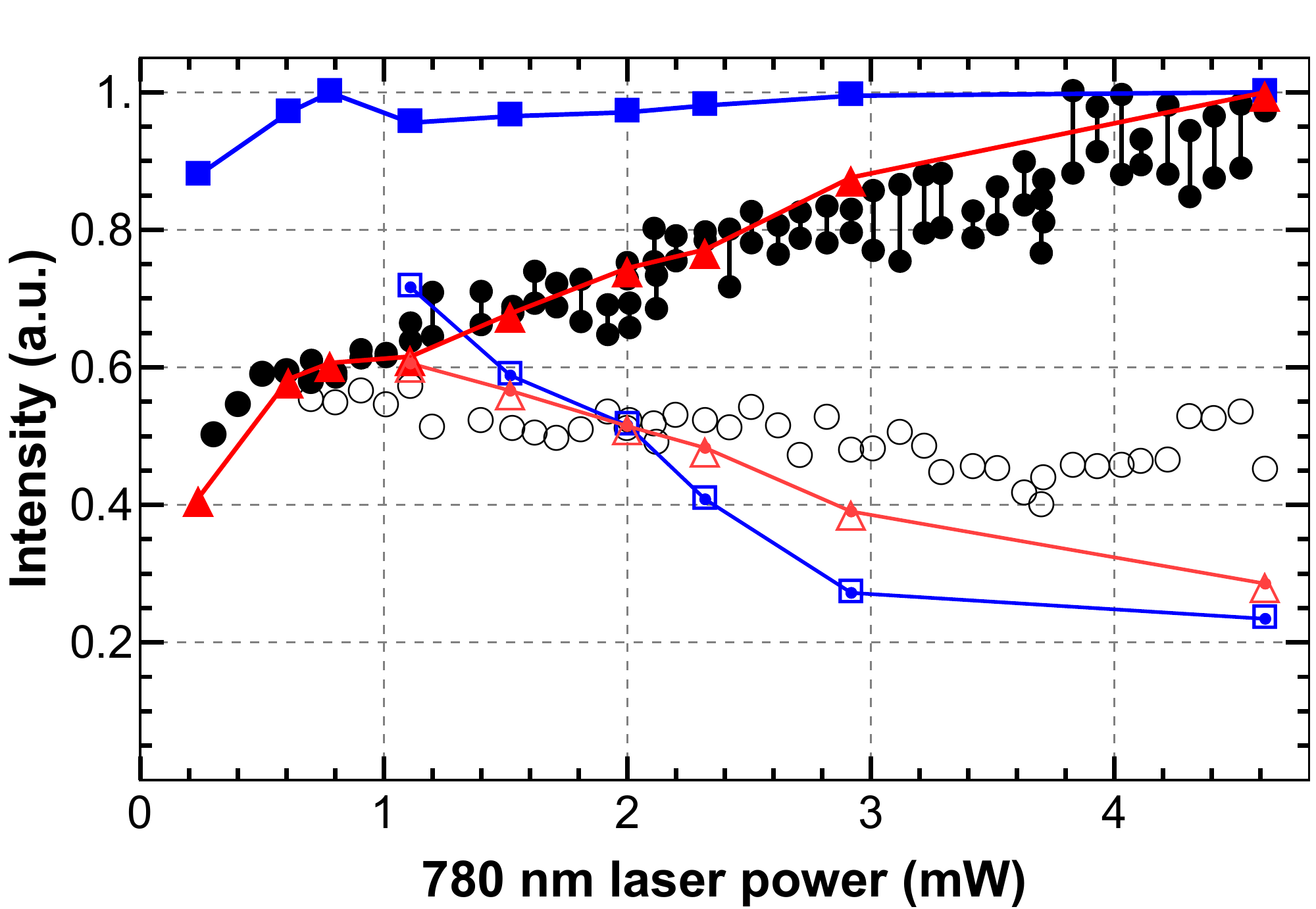}
    \caption{\label{fig:minmax} Comparison of the maximum intensity of the Autler-Townes doublets (filled markers) and the minima formed between the peaks (empty markers) for the $F_2 =2\to F_3 =2$ transition component. Empty and filled circles correspond to the experimental data. As a result of variations of the specific frequency to which the $\unit{780}{\nano\metre}$ laser is locked with respect to the centre of the transition resonance at the moment of recording each one of the measured spectra, there are imbalances on the heights of the maximum intensities of the AT doublets. The points joined by vertical lines indicate these two maximum intensities in each of the recorded spectra. The 3-level and 4-level theoretical maxima and minima are shown respectively with square and triangular markers.  }
\end{figure}

The graphs in Fig.~\ref{fig:exp} show that the growth of the resonance peaks and the evolution of the Autler-Townes spitting as a function of the power of the $\unit{780}{\nano\metre}$ coupling laser observed in the experimental data is more closely reproduced by the 4-level model than by the simpler 3-level model. 
This is confirmed in Fig.~\ref{fig:minmax} where the evolution of the resonance maximum before the ATS appears and of the maxima and central minimum of the AT doublet in the most prominent transition component ($F_2 =2\to F_3 =2$) are plotted against the power of the $\unit{780}{\nano\metre}$ coupling light. The simulated spectra obtained by means of the 3-level model present a rapid growth, reaching saturation below $\unit{0.8}{\milli\watt}$, and showing the first evidence of the ATS at around  $\unit{1}{\milli\watt}$, value after which the maxima of the AT doublets reach a plateau.
Also in this simple model, the value of the minimum in the middle of the AT peaks drops quickly to a fraction below $75\%$ of the value of the AT maxima for a power of $\unit{3}{\milli\watt}$ and drops to nearly $80\%$ at the far end of the measured range. 
The experimental spectra on the other hand reach a transitory saturation between $\unit{0.5}{\milli\watt}$ and $\unit{1}{\milli\watt}$, where the AT doublet begins to be clearly discernible in the shape of the spectra. At higher powers of the coupling light, the maximum values of the measured doublets keep increasing without showing any evidence of saturating again. This is in contrast to the behaviour of the central minima, which show hardly any change from the onset of the splitting at a fraction $\lesssim 60\%$ of the maximum value reached by the fluorescence at the highest measured power, and seem to saturate at a value slightly below $50\%$.  
The overall behavior of the experimental spectra is better reproduced by the calculations generated with the aid of the 4-level model.

\section{Conclusions.}\label{S:Conclusions}


In this paper, it has been shown that the highly non-perturbing nature of an electric dipole forbidden transition, the electric quadrupole interaction term in the case hereby presented, provides an ideal probe for performing an in-depth investigation of the dynamics of an atomic system interacting with radiation fields.
We demonstrate this method by probing the ATS that relies on the utilization of a forbidden transition and allows experimental and theoretical studies under nontrivial conditions. 
In this system, the usage of a velocity-selective scheme based on counter-propagating probe and control lasers establishes limits to the Doppler contributions caused by working with a warm atomic sample.
In our method, the proper selection of the polarization of the control and probe lasers simplified the hyperfine structure manifested in the ATS. This allowed a simple description of the experimental results requiring only a few physical parameters with a clear significance.

The theoretical description involved three- and four-level systems including one forbidden electric dipole transition, which makes it significantly different from the standard studies based on E1 selection rules. The three-level ladder configuration admits the possibility of a parametric up-conversion process: the absorption of two photons with a given frequency can lead to the spontaneous emission of a photon with higher frequency. The four-level model considers an extra state to represent the joint effects of the alternative decay routes. This methodology enables simplified and compact numerical simulations, an efficient time-dependent analysis of the Bloch equations, and yields rather realistic predictions that closely reproduce the observations.

We also derived a simple expression for evaluating finite temperature  effects and the laser beam bandwidths. The formalism here presented circumvents solving Bloch equations incorporating Doppler detunings followed by velocity and, spectral beam averages by the simple \emph{a posteriori} averaging of the solutions of the Bloch equations as a function of the detunings $\delta_{12}$ and $\delta_{23}$ using effective temperature-dependent frequency distributions. An interesting result derived from this formalism is that the decay fluorescence not only exhibits ATS as a function of the detuning of the probe laser, but also does it as a function of the detuning of the control laser in a counter-propagating configuration at room temperature. Furthermore, we showed that via this scheme one can perform a direct study of the ATS broadening for non-counter-propagating beams. In this way, we could estimate the time required to achieve a steady state for the density matrix both in the Bloch and Maxwell-Boltzmann-Bloch schemes, and compare this time with the average transit time of the atoms within the laser beams. With this analysis we concluded that, under the experimental conditions of the measurements presented in this paper, the atoms transit times within the laser beams were comparable to those required to obtain a steady state.
This prevents reaching saturation on the transition induced by the control laser at beam powers that would yield it for atoms at rest.
%


To finalize, note that the use of cooling and trapping techniques would eliminate the Doppler and transit time issues considered in this work. Thus, the method proposed here would be readily applicable in such systems as a minimal perturbing method to implement and evaluate the efficiency of protocols for the preparation of atomic states.




\section{Appendix A}

\begin{figure}
\includegraphics[width = 0.35\textwidth]{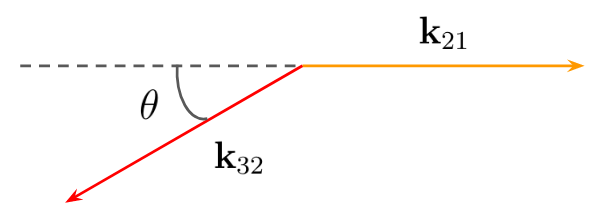}%
\caption{Wave vectors of the control $\vec k_{21}$ and probe $\vec k_{32}$ lasers.\label{fig:lasers}}
\end{figure}

For control and probe lasers in a non-counter-propagating configuration, the distribution that includes the Doppler and lasers detunings differs from that given by
Eq.~(\ref{eq:Dwidth}).
For the geometry described in Fig.~\ref{fig:lasers}, a direct calculation shows that
\begin{equation}
     \tilde \rho_{ij}^{(\sigma_{lm},T)}(\delta_{21}^{(0)},\delta_{32}^{(0)}) =\Big( \frac{m\upsilon^{||2}_D}{k_{\mathrm{B}}T}\Big)^{1/2}\Big( \frac{m\upsilon^{\bot 2}_D}{k_{\mathrm{B}}T}\Big)^{1/2}
 \int d\delta_{21}\int d\delta_{32} e^{-\kappa_{||}(\delta_{21},\delta_{32})- \kappa_{\bot}(\delta_{21},\delta_{32})}  \rho_{ij}(\delta_{21} ,\delta_{32}) 
\end{equation}
where
\begin{eqnarray}
    \frac{1}{\upsilon^{||2}_D} &=&\Big(\frac{m}{k_{\mathrm{B}} T} + \frac{\vert k_{21}\vert ^2}{\sigma_{21}^2}  
    + \frac{\vert k_{32}\vert ^2\cos^2\theta}{\sigma_{32}^2}\Big)\label{eq:vDa}\\
    \kappa_{||}(\delta_{21},\delta_{32})&=& \Big(\frac{\delta_{21}- \delta_{21}^{(0)}}{\sqrt{2}\tilde \sigma_{21}} \Big)^2
     +\Big(\frac{\delta_{32}- \delta_{32}^{(0)}}{\sqrt{2}\tilde \sigma_{32}} \Big)^2\nonumber\\
     &+& \Big(\frac{(\delta_{21}- \delta_{21}^{(0)})}{\sigma_{21}}\frac{\vert k_{21}\vert\upsilon_D}{ \sigma_{21}}\Big)
     \Big(\frac{(\delta_{32}- \delta_{32}^{(0)})}{\sigma_{32}}\frac{\vert k_{32}\vert\cos\theta\upsilon^{||}_D}{ \sigma_{32}}\Big)\label{eq:kappaa}\\
     \tilde\sigma_{21}^2 &=& \frac{1}{1 - \vert k_{21}\vert^2\upsilon_D^{||2}/\sigma_{21}^2} \sigma_{21}^2\label{eq:sigma21a} \\
    \tilde\sigma_{32}^2 &=& \frac{1}{1 - \vert k_{32}\cos\theta\vert^2\upsilon^{|| 2}_D/\sigma_{32}^2}\sigma_{32}^2\label{eq:sigma32a} 
\end{eqnarray}
and
\begin{eqnarray}
    \frac{1}{\upsilon^{\bot 2}_D} &=&\Big(\frac{m}{k_{\mathrm{B}} T} + \frac{\vert k_{32}\sin\theta\vert ^2}{\sigma_{32}^2} \Big)\label{eq:vDab}\\
    \kappa_{\bot}(\delta_{21},\delta_{32})&=& \Big(\frac{\delta_{32}-\delta^{(0)}_{32}}{\sqrt{2} \sigma_{32}^2} + \frac{\delta_{21}-\delta^{(0)}_{21}}{ \sqrt{2}\sigma_{21}^2}\frac{\vert k_{21}\vert
    \vert k_{32}\cos\theta\vert\upsilon^{||2}_D }{\sigma_{32}^2} \Big)^2\upsilon^{\bot 2}_Dk_{32}^2\sin^2\theta
\end{eqnarray}

\begin{acknowledgments}
We thank J. Rangel for his help in the construction of the diode laser. This work was supported by DGAPA-UNAM, M\'exico, under projects PAPIIT Nos. IN116309, IN110812, and IA101012, and by CONACyT, M\'exico, under project No. 44986, LN-LANMAC-CTIC-2019 and PIIF-{\it Correlaciones cu\'anticas: teor\'{\i}a y experimento}-2019. L.M. Hoyos-Campo thanks UNAM-DGAPA and Conacyt for the postdoctoral fellowship.
\end{acknowledgments}

%


\end{document}